\documentclass[11pt]{article}
\usepackage[T1]{fontenc}
\usepackage{lmodern,amsmath,amssymb,amsthm,mathtools}
\usepackage[a4paper,margin=25mm]{geometry}
\usepackage{enumitem,microtype,graphicx,tikz}
\usetikzlibrary{arrows.meta,calc,positioning,fit}
\usepackage[numbers,sort&compress]{natbib}
\usepackage[colorlinks=true,linkcolor=blue,citecolor=blue,urlcolor=blue]{hyperref}
\newtheorem{theorem}{Theorem}[section]
\newtheorem{lemma}[theorem]{Lemma}

\newtheorem{corollary}[theorem]{Corollary}
\newtheorem{remark}[theorem]{Remark}
\numberwithin{equation}{section}
\newcommand{\E}{\mathbb E}
\newcommand{\Prb}{\mathbb P}
\newcommand{\ceil}[1]{\left\lceil #1\right\rceil}
\newcommand{\floor}[1]{\left\lfloor #1\right\rfloor}
\newcommand{\col}{\operatorname{col}}

\newcommand{\mono}{\mathrm{mono}}
\newcommand{\highdeg}{\mathrm{highdeg}}
\newcommand{\lowdeg}{\mathrm{lowdeg}}
\newsavebox{\vertexflowchartbox}

\title{New bounds for online coloring three-colorable graphs}
\author{Yaqiao Li\footnote{Shenzhen University of Advanced Technology, Shenzhen, China, liyaqiao@suat-sz.edu.cn}}
\date{\today}

\begin{document}
\maketitle
\begin{abstract}
We prove new lower and upper bounds for online coloring of three-colorable graphs, improving classical bounds of Vishwanathan [FOCS'1990] that have stood for more than three decades. For the lower bound against deterministic algorithms, we improve Vishwanathan's $\Omega(\log^2n)$ to \(\exp\!\bigl(\Omega((\log n)^{1/3}(\log\log n)^{2/3})\bigr)\). For the upper bound of randomized algorithms, we give a randomized algorithm that achieves $O(\sqrt{n})$ improving Vishwanathan's $O(\sqrt{n \log n})$. 

Technically, for the lower bound construction, we introduce the language of $t$-fold coloring for simpler description of recursive construction, and the notion of post-degree for future control, and use the flexibility of two-sided colors of bipartite graph, constructed using the independent set amplification method of Vishwanathan, as building blocks to ensure three-colorability. The randomized algorithm integrates, in a new way,  some of the ideas in offline coloring of Wigderson and Blum with ideas of Vishwanathan in the online setting, such as the degree control from Wigderson, the offline-monochromatic class from Blum, and the anchor-sampling from Vishwanathan and the CBIP algorithm by Lov{\'a}sz, Saks and Trotter.

\end{abstract}

\section{Introduction}\label{sec:introduction}

Three-colorable graphs are computationally complicated objects: up to today, there is still no efficient algorithm to decide whether a given graph is three-colorable, in fact this is one formulation of the P vs NP problem. Besides computation, another perspective to study them is via information: how limited information inhibits the coloring even the computational time is allowed to be super-polynomial. 

One model of studying limited information is the online graph  vertex-coloring model.  An online vertex-coloring algorithm receives vertices one at a time,
together with their edges to earlier vertices. It must assign each
vertex a color immediately and irrevocably. 
For a graph \(G\), an order \(\pi\), and an algorithm \(\mathcal A\),
write \(C_{\mathcal A}(G,\pi)\) for the number of colors used. In 1990, Vishwanathan \cite{Vishwanathan1990} proved that every deterministic online algorithm requires  $\Omega(\log^2 n)$ colors to color three-colorable graphs on $n$ vertices, and gave a randomized algorithm that uses $O(\sqrt{n\log n})$ colors in expectation. These two bounds stood for more than three decades. This work improve both.

\begin{theorem}[Deterministic lower bound]\label{thm:deterministic}
There is an absolute constant \(c>0\) such that, for every sufficiently
large \(n\) and every deterministic online coloring algorithm
\(\mathcal A\), some ordered three-colorable graph \(G\) on \(n\)
vertices satisfies
\[
 C_{\mathcal A}(G,\pi)\ge
 2^{c(\log n)^{1/3}(\log\log n)^{2/3}}.
\]
\end{theorem}

\begin{theorem}[Randomized upper bound]\label{thm:randomized}
There is a randomized online algorithm
\(\mathcal A\) such that, for every fixed ordered three-colorable graph
with at most \(n\) vertices,
\[
 \E C_{\mathcal A}(G,\pi)=O(\sqrt n).
\]
The algorithm can be
implemented in expected polynomial time.
\end{theorem}

\begin{remark}  \label{rmk:after_thm}
    1. By a technique introduced in Lovasz et al \cite{LST1989}, one can assume online algorithms know the number of vertices, with only $\Theta(1)$ multiplicative difference in the colors used. So we assume this when it is convenient. In particular, this implies the lower bound holds even if algorithms know $n$. 
    
    2. The lower bound Theorem \ref{thm:deterministic} can be extended to work against randomized algorithms, and the randomized algorithm Theorem \ref{thm:randomized} can be adapted to give a deterministic algorithm in the random order model with the same bound: these are provided in the appendix. 

    3. As in Vishwanathan \cite{Vishwanathan1990}, the two bounds we obtained could be generalized to get new bounds for online coloring on $k$-colorable graphs for $k\ge 4$. We did not try to do that.
\end{remark}

\begingroup
\renewcommand{\O}{O}
\subsection{Closely related work}\label{sec:related}

\paragraph{Offline coloring three-colorable graphs: Wigderson and Blum.}
Wigderson's \(O(\sqrt n)\)-color algorithm \cite{Wigderson1983}
used two facts: the neighborhood of a vertex  in a three-colorable graph is
bipartite, and a graph of maximum degree \(\Delta\) admits a greedy
\((\Delta+1)\)-coloring. Wigderson's algorithm then proceeded by checking vertex degree: for degree at least $\sqrt{n}$, it two-colored its neighborhood of  at least \(\sqrt n\) vertices and removed the neighborhood. There were at most
\(\sqrt n\) such removals. The remaining graph had degree below
\(\sqrt n\) and was colored greedily.

Blum \cite{Blum1994} obtained
\(O(n^{3/8}\operatorname{polylog}n)\) colors through several forms of
combinatorial progress that he initiated and became a foundational framework for subsequent studies. One was to certify vertices that have the same
color in \emph{every} proper three-coloring and contract them, reducing the graph
size. A collection of vertices certified in this way can be compressed into a
single vertex.  We call such a collection an \emph{offline-monochromatic class}. For example, if $uv$ is an edge, then all vertices incident to both $u$ and $v$ enjoy this property. 
This was one component of Blum's framework, alongside independent-set
and neighborhood arguments. Later offline algorithms combined
combinatorial progress with semidefinite methods; examples included
Blum--Karger \cite{BlumKarger1997} and
Kawarabayashi--Thorup--Yoneda \cite{KawarabayashiThorupYoneda2024}. 

Some parts of Wigderson's method were adapted to the online setting by Vishwanathan \cite{Vishwanathan1990}. We use that too. Furthermore, we adapt Blum's offline-monochromatic class compression into the online setting, suitably combined with a degree-control.

\paragraph{Online coloring bipartite graphs: FirstFit and CBIP.}
The algorithm FirstFit assigns the least positive integer absent from the earlier
neighbors' colors. Gy\'arf\'as and Lehel \cite{GyarfasLehel1988} studied FirstFit online coloring on restricted graph classes,  in particular, they gave an online $n$-vertex forest (hence, bipartite) construction  forcing $\Omega(\log n)$ colors on every deterministic algorithm. 

It is not hard to see that FirstFit can use up to $\Theta(n)$ colors on bipartite graphs of $n$ vertices, such as on crown graphs, see Ye et al. \cite{ye2026firstfit}. However, a useful feature of FirstFit is that if it uses color $k$ on an online vertex $v$, it is guaranteed that there are at least $k-1$ vertices arrived before $v$ that are all connected to $v$ and exhausted the colors $\{1,\ldots, k-1\}$. 

Lov\'asz, Saks, and Trotter \cite{LST1989} gave an online coloring algorithm, which became known as CBIP, using at most \(2\log n+O(1)\) colors for bipartite graphs on $n$ vertices.
For the arriving vertex, the algorithm CBIP
took the least color absent from the opposite side of its current
connected component.  Bianchi et al.\ \cite{BianchiEtAl2014} gave a component-growth
proof of this upper bound and improved the bipartite lower bound.
Gutowski et al. \cite{GutowskiEtAl2014} proved the
matching lower bound \(2\log n-O(1)\). Their construction explicitly
tracked \emph{two-sided colors}, namely colors appearing on both
sides of a bipartite component.
Li et al. \cite{LiNarayanPankratov2022} further gave a refined analysis of CBIP and two-sided colors, obtaining a \(2\kappa\) bound when at most \(\kappa\) connected components are present at any time. 

In our lower-bound construction for three-colorable graphs, we construct bipartite graphs whose colors of interest are two-sided, and use these  bipartite graphs as building blocks to construct larger graphs whose three-colorability is ensured by the  flexibility of two-sided colors of the bipartite graphs, while forcing new online colors.  In our randomized online algorithm design, we take advantage of the feature of FirstFit discussed above and the fact that CBIP uses $\Theta(\log n)$ colors on bipartite graphs: both were already used by Vishwanathan \cite{Vishwanathan1990}. 

\paragraph{Vishwanathan's online lower and upper   bounds on three-colorable graphs.}
Vishwanathan \cite{Vishwanathan1990} gave  an
\(\Omega(\log^2 n)\) lower bound  and an
\(O(\sqrt{n\log n})\) randomized upper bound for three-colorable graphs; see also
Kierstead's survey \cite{KiersteadSurvey1998}.

For the lower bound, observe that it is equivalent to construct an independent set (of a three-colorable graph) that is forced to use lots of colors: this is because the whole graph can be decomposed into at most three independent sets. 
The core of Vishwanathan's lower bound construction might be called \emph{independent set amplification}, as illustrated in Figure ~\ref{fig:vishwanathan-lower}.
Suppose a construction produces an independent set \(A\) with at
least \(k\) online colors. Construct a disjoint copy with such a set \(B\).
Next construct a bipartite graph $H$, joining all its vertices to \(A\).
One of two parts of $H$, say \(C\), has \(\Omega(\log m)\)
colors on an \(m\)-vertex bipartite construction. Note that colors used on $C$ is necessarily disjoint from colors used on
\(A\). A simple calculation shows that either the independent set \(A\cup B\) or  
\(B\cup C\) has at least \(k+\Omega(\log m)\) colors. This amplification process can then be repeated within certain range of the related parameters, forcing at least $\Omega(\log^2 n)$ colors. This technique is also used in our lower bound construction.

\begin{figure}[tbp]
\centering
\begingroup
\definecolor{vishblue}{RGB}{31,72,105}
\definecolor{vishpale}{RGB}{239,245,249}
\begin{tikzpicture}[x=1cm,y=1cm,font=\small,
  vishvertex/.style={circle,draw=vishblue,fill=vishblue,
    minimum size=.11cm,inner sep=0pt},
  vishoutline/.style={draw=vishblue,line width=.7pt,fill=vishpale},
  vishedge/.style={draw=vishblue!75,line width=.55pt},
  vishjoin/.style={draw=vishblue,line width=1.35pt}]

  \draw[vishedge] (-3.46,1.13)--(-4.53,-.13);
  \draw[vishedge] (-3.23,1.07)--(-4.27,-.11);
  \draw[vishedge] (-3.04,1.07)--(-2.12,-.11);
  \draw[vishedge] (-2.82,1.14)--(-1.87,-.12);
  \draw[vishedge] (-3.84,-.19)--(-2.47,-.19);
  \draw[vishedge] (-3.83,-.40)--(-2.48,-.40);

  \draw[vishedge] (2.84,1.13)--(1.77,-.13);
  \draw[vishedge] (3.07,1.07)--(2.03,-.11);
  \draw[vishedge] (3.26,1.07)--(4.18,-.11);
  \draw[vishedge] (3.48,1.14)--(4.43,-.12);
  \draw[vishedge] (2.46,-.19)--(3.83,-.19);
  \draw[vishedge] (2.47,-.40)--(3.82,-.40);

  \draw[vishjoin] (-2.73,1.57)--(-1.90,2.90);

  \foreach \xx/\yy in {-3.15/1.35,-4.35/-.30,-1.95/-.30,
                        3.15/1.35,1.95/-.30,4.35/-.30}
    \path[vishoutline] (\xx,\yy) ellipse[x radius=.69cm,y radius=.39cm];
  \foreach \xx in {-3.49,-3.15,-2.81,2.81,3.15,3.49}
    \node[vishvertex] at (\xx,1.35) {};
  \foreach \xx in {-4.57,-4.13,-2.17,-1.73,1.73,2.17,4.13,4.57}
    \node[vishvertex] at (\xx,-.30) {};
  \node[above] at (-3.15,1.76) {\(A\)};
  \node[above] at (3.15,1.76) {\(B\)};
  \node[font=\small\bfseries] at (-3.15,-.92) {\(G\)};
  \node[font=\small\bfseries] at (3.15,-.92) {\(G'\)};

  \path[draw=vishblue,line width=.7pt,rounded corners=2pt,fill=white]
    (-2.10,2.90) rectangle (1.10,5.35);
  \node[right,font=\small\bfseries] at (1.20,3.98) {\(H\)};
  \draw[vishedge] (-1.28,4.46)--(-1.12,3.49);
  \draw[vishedge] (-.76,4.46)--(-1.12,3.49);
  \draw[vishedge] (-.76,4.46)--(-.50,3.49);
  \draw[vishedge] (-.24,4.46)--(.12,3.49);
  \draw[vishedge] (.28,4.46)--(.12,3.49);
  \path[vishoutline] (-.50,4.46) ellipse[x radius=1.24cm,y radius=.34cm];
  \path[vishoutline] (-.50,3.49) ellipse[x radius=1.24cm,y radius=.34cm];
  \node[inner sep=1pt] at (-.50,5.03) {\(C\)};
  \foreach \xx in {-1.28,-.76,-.24,.28}
    \node[vishvertex] at (\xx,4.46) {};
  \foreach \xx in {-1.12,-.50,.12}
    \node[vishvertex] at (\xx,3.49) {};

  \node[font=\footnotesize] at (0,-1.52)
    {Arrival order: \(G\), then \(G'\), then \(H\). };
\end{tikzpicture}
\endgroup
\caption{The independent set amplification in Vishwanathan's lower-bound construction: either $A\cup B$ or $B\cup C$ forces $\Omega(\log m)$ new online colors  if $|V(H)| = m$.}
\label{fig:vishwanathan-lower}
\end{figure}
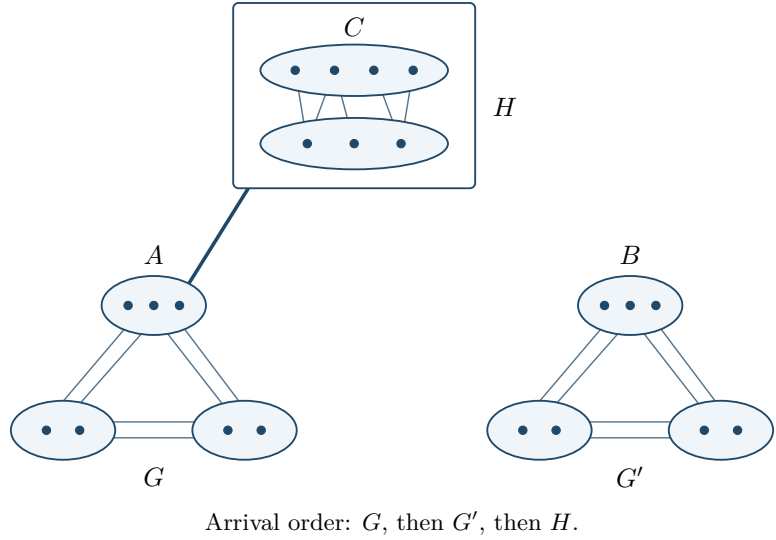

For the upper bound, it could be helpful to think of it as a randomized online version of Wigderson's offline algorithm, see Figure~\ref{fig:vishwanathan-upper}  for an illustration. The algorithm reserved a separate palette
of \(c\) FirstFit colors. At the start it chose an index \(J\) uniformly
from \([c]\). For each arriving online vertex the algorithm tried to color it using FirstFit with at most $c$ colors. In case when an online vertex $v$ needed to use color $c+1$ by FirstFit, the feature of FirstFit guaranteed a neighbor $u$ of $v$ in the color class $J$. The vertex $u$ was then used as an \emph{anchor} and $v$ was colored by CBIP in the bipartite neighborhood of $u$, using at most $O(\log n)$ colors. Each anchor $u$ used a separate palette of at most $O(\log n)$ colors. Since the expected size of
the vertices in $J$ was at most \(n/c\), the total colors was at most
\[
 c+O\bigl((n/c)\log n\bigr)
   =O(\sqrt{n\log n})
 \quad\text{for }c=\Theta(\sqrt{n\log n}).
\]
Note that the expected number of anchors was $n/c = O(\sqrt{n/\log n})$. 
Like Wigderson's algorithm, this exploited bipartite neighborhoods;
the logarithmic online cost replaced offline two-coloring.

\begin{figure}[tbp]
\centering
\begingroup
\definecolor{vnavy}{RGB}{31,72,105}
\definecolor{vpale}{RGB}{239,245,249}
\resizebox{\linewidth}{!}{%
\begin{tikzpicture}[x=1cm,y=1cm,font=\small,
  vcollection/.style={draw=vnavy,line width=.6pt,rounded corners=2pt,
    fill=white,inner sep=4pt},
  vvertex/.style={circle,draw=vnavy,line width=.6pt,fill=white,
    inner sep=0pt,minimum size=1.9mm},
  vnamed/.style={vvertex,minimum size=5mm,inner sep=1pt},
  vedge/.style={draw=vnavy,line width=.55pt}]

  \node[font=\small\bfseries,align=center,text width=7.6cm]
    at (-3.9,3.8) {(a) FirstFit with \(c\) colors};
  \node[font=\footnotesize,align=center,text width=7.6cm]
    at (-3.9,3.25)
      {Choose \(J\) uniformly from \([c]\) at the start.};

  \foreach \ii/\xx in {1/-7.1,2/-5.65,c/-.7} {
    \node[vcollection,minimum width=.72cm,minimum height=2.8cm]
      at (\xx,1) {};
    \node[font=\small] at (\xx,2.7) {\(\ii\)};
  }
  \node[vcollection,fill=vpale,line width=1pt,
    minimum width=.72cm,minimum height=2.8cm] at (-2.75,1) {};
  \node[font=\small\bfseries] at (-2.75,2.7) {\(J\)};
  \foreach \xx in {-4.2,-1.7} {
    \node at (\xx,2.7) {\(\cdots\)};
    \node at (\xx,1) {\(\cdots\)};
  }

  \foreach \xx in {-7.1,-5.65,-2.75,-.7}
    \draw[vedge] (-4,-1.9)--(\xx,-.16);
  \foreach \xx in {-7.1,-5.65,-2.75,-.7} {
    \foreach \yy in {2.02,1.29,.56}
      \node[vvertex] at (\xx,\yy) {};
  }
  \foreach \xx in {-7.1,-5.65,-.7}
    \node[vvertex] at (\xx,-.16) {};
  \node[vnamed,line width=1pt,minimum size=4mm]
    (ffanchor) at (-2.75,-.16) {\(u\)};
  \node[vnamed] at (-4,-1.9) {\(v\)};
  \node[align=center,text width=7.4cm,font=\footnotesize] at (-3.9,-2.9)
    {If attempted FirstFit color of \(v\) exceeds \(c\), then choose an earlier neighbor \(u\) in the selected class \(J\).\\
     Every vertex in \(J\) is an anchor.};

  \node[font=\small\bfseries,align=center,text width=6.8cm]
    at (4.75,3.8) {(b) CBIP in the anchor's neighborhood};
  \node[vcollection,minimum width=4.9cm,minimum height=3.1cm]
    (neighborhood) at (4.75,.1) {};
  \node[vnamed,fill=vpale,line width=1pt]
    (cbipanchor) at (4.75,2.6) {\(u\)};
  \node[font=\footnotesize,anchor=west] at (5.13,2.6) {anchor};
  \draw[vedge,line width=1.35pt]
    (cbipanchor.south)--(neighborhood.north);

  \draw[vedge] (3.1,1)--(6.4,1);
  \draw[vedge] (3.1,1)--(6.4,.25);
  \draw[vedge] (3.1,.25)--(6.4,1);
  \draw[vedge] (3.1,.25)--(6.4,-.5);
  \draw[vedge] (3.1,-.5)--(6.4,.25);
  \draw[vedge] (3.1,-.5)--(6.4,-.5);

  \foreach \yy in {1,.25,-.5}
    \node[vvertex,minimum size=2.6mm] at (3.1,\yy) {};
  \foreach \yy in {1,.25}
    \node[vvertex,minimum size=2.6mm] at (6.4,\yy) {};
  \node[vnamed] at (6.4,-.5) {\(v\)};
  \node at (4.75,-1.85) {\(G[N(u)]\) is bipartite};
  \node[align=center,text width=6.8cm,font=\footnotesize] at (4.75,-2.9)
    {Since \(v\in N(u)\), use \(u\)'s own CBIP palette,
     disjoint from the FirstFit palette and other anchors' palettes.};
\end{tikzpicture}%
}
\endgroup
\caption{Vishwanathan's randomized algorithm.}
\label{fig:vishwanathan-upper}
\end{figure}
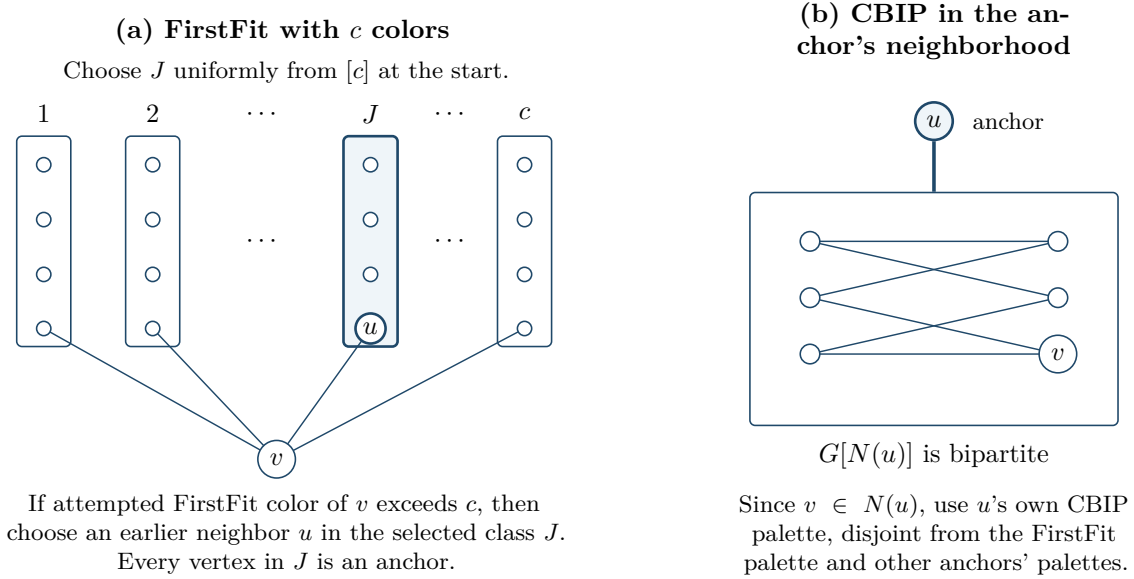

For comparison, on the deterministic upper-bound side, Kierstead
\cite{Kierstead1998} gave
\(O(n^{2/3}(\log n)^{1/3})\) colors for three-colorable graphs. 
For $k$-colorable graphs with $k \ge 4$, Kierstead
\cite{Kierstead1998} also obtained upper bounds which were improved in the recent breakthrough  Kawarabayashi, Yoneda, and Yoneda
\cite{KawarabayashiYonedaYoneda2025}.

\subsection{Conceptual and technical contributions}\label{sec:contributions}

\paragraph{Deterministic lower bound.}

Our lower bound construction uses the similar independent set amplification as in Vishwanathan. A \emph{conceptually} new way of doing the construction is that we use a $t$-fold online coloring model: each vertex is assigned a set of $t$ distinct colors, and distinct vertices use disjoint color sets. The standard online coloring model corresponds to $t=1$. Suppose our goal is to force $K$ colors, we can trivially force it using a single vertex in the $K$-fold coloring model, and then our task is to ask whether it is possible to force $K$ colors in the standard $1$-fold coloring and using how many vertices. This suggests interpolation and simulation: if one can force $K$ colors using $q$-fold coloring and $N$ vertices, given $t<q$, can you simulate it (i.e., also force $K$ colors) using $t$-fold coloring and $N'$ vertices for some $N'$ as a function of $K, N,q,t$? Clearly, such a simulation lemma can then be used to build a lower bound construction. 

Hence, fixing $K$, the goal is to upper bound $N'$ as a function of $K, N,q,t$. It turns out that a certain control of the \emph{future} is helpful. We use \emph{post-degree} to control the future: the post-degree of an online vertex is the number of its future neighbors. To the best knowledge of the author, making use of the post-degree in online coloring lower bound construction seems to be new.

Technically, the online three-colorable graphs are constructed using bipartite graphs as building blocks. In particular, for the simulation with $t<q$, given a construction of the $q$-fold coloring, each online vertex in the $q$-coloring model will be simulated by a collection of disjoint copies of certain bipartite graphs in the $t$-fold coloring model. Each such bipartite graph has sufficiently many two-sided colors (which are produced via the independent set amplification), which allows to simulate the structure of the three-colorable graph in the $q$-fold coloring model and force a set of $q$ new colors. The post-degree parameter in the $q$-fold coloring gives us a handle to control both the number of vertices and the post-degree of the construction in the $t$-fold coloring.

\paragraph{Randomized upper bound.}
We combine Vishwanathan's anchor-sampling  with an online form of Blum's offline-monochromatic class compression and Wigderson's degree control, in a new way. 

Our randomized algorithm also uses the idea of anchor-sampling, but in expectation uses $O(\sqrt{n})$ anchors instead of $O(\sqrt{n/\log n})$. The main departure from Vishwanathan's algorithm is that we only allow CBIP  to use  $O(1)$ colors to color an anchor's bipartite neighborhood, and send vertices exceeding this bound to recursive processing. It can be shown that such vertices necessarily form an offline-monochromatic class which allows for a similar effect of compression as in Blum's offline coloring. For vertices with no anchored neighbor, we then process them by their degree, and similarly as in Wigderson simply color the low-degree vertices by (a randomized version of) FirstFit, and send the high-degree vertices to another recursive stream. For the low-degree vertices, we also only allow FirstFit to use $O(1)$ colors, and for vertices that exceed this bound we again send them to another recursive stream. By properly designing the anchor-sampling, the degree threshold, and those disjoint palettes' sizes, each of the three recursions is guaranteed to receive only a constant fraction of vertices, which provides an effect of compression that leads to $\O(\sqrt{n})$ colors in expectation. As an example, the anchor-sampling and the degree control must work together to ensure that  high degree vertices without anchored neighbor are not too many.

\endgroup



\section{A lower bound for deterministic algorithms}\label{sec:lower}

We use  \emph{\(t\)-fold coloring} to describe the recursive construction of the graph. In a $t$-fold online coloring model, a vertex receives a set of \(t\)
distinct online colors, and adjacent vertices receive disjoint sets. We fix the target number \(K\) of online colors throughout the construction. The auxiliary \(t\)-fold model below lets us pass from one vertex in the \(K\)-fold model to an ordinary coloring construction
by a sequence of simulations, without changing this target.

We start with two simple combinatorial lemma that will be used a couple of times later. 

\begin{lemma}\label{lem:union}
Let \(P,Q,X\) be finite sets with \(|P|,|Q|\ge a\), \(|X|=t\),
and \(X\cap P=\varnothing\). Put \(s=\ceil{t/2}\). Then
\begin{equation}
 \max\{|P\cup Q|,\ |Q\cup X|\}\ge a+s.
 \label{eq:union}
\end{equation}
\end{lemma}
\begin{proof}
If \(|P\cup Q|\ge a+s\), there is nothing to prove. Otherwise, 
 $|Q\setminus P|\le a+s-1-|P|\le s-1$.
Since \(X\) avoids \(P\), its intersection with \(Q\) has size at most
\(s-1\). Consequently,
 $|Q\cup X|\ge a+t-s+1\ge a+s$.
\end{proof}

\begin{lemma}\label{lem:degree-count}
Let \(G\) be a bipartite graph with sides \(A,B\), where
\(|A|=a\ge1\) and \(|B|=b\). Suppose that every vertex of \(A\) has
degree at least the integer \(d\ge1\). For every integer \(1\le i\le a\),
\begin{equation}
 \bigl|\{u\in B:\deg(u)\ge i\}\bigr|
 \ge
 \ceil{\frac{ad-b(i-1)}{a-i+1}}.
 \label{eq:degree-count}
\end{equation}
In particular, if \(a>b\), at least \(d\) vertices of \(B\) have
degree at least two.
\end{lemma}
\begin{proof}
Write \(b_i=|\{u\in B:\deg(u)\ge i\}|\) and let \(e=|E(G)|\). We use double counting of the edges.
Counting edges at \(A\) gives \(e\ge ad\). At \(B\), each of the
\(b_i\) vertices has degree at most \(a\), while all the others have
degree at most \(i-1\). Hence
\[
 ad\le e\le ab_i+(b-b_i)(i-1)
       =b(i-1)+(a-i+1)b_i.
\]
Rearranging proves the claim. The in particular part can be directly verified with \(i=2\).
\end{proof}

\subsection{Bipartite graphs forcing many two-sided colors}
\label{sec:bipartite-lower}

The goal of this part is to use the independent set amplification technique to give a construction of a bipartite graph forcing many two-sided colors in every $t$-fold online coloring. These bipartite graphs will later be used as basic building blocks for simulation, or equivalently, in the recursive construction of three-colorable graphs.

For a colored vertex set \(S\), let \(\col(S)\) be the union of its assigned online colors.

We first generalize the forest bipartite construction to the $t$-fold coloring model, integrating the independent set amplification. For convenience, for bipartite graphs with a bipartition, we use side to denote one part of the bipartition.

\begin{lemma}\label{lem:bipartite-independent}
For integers \(t\ge1\) and \(j\ge0\), and every proper \(t\)-fold coloring algorithm, there exists an adaptive bipartite graph construction using at most \(2^{j+1}-1\) vertices such that one of its two sides, call it \(I\), satisfying
 $|\col(I)|\ge t+j\ceil{t/2}$.
\end{lemma}
\begin{proof}
The construction will be inductive. 
At \(j=0\), present one vertex and take it as \(I\). For \(j\ge1\),
present two disjoint copies of the level-\((j-1)\) construction, with
designated sides \(I_1,I_2\), respectively. Write
\[
 P=\col(I_1),\qquad Q=\col(I_2),\qquad
 a=t+(j-1)s,\qquad s=\ceil{t/2}.
\]
By induction, \(|P|, |Q| \ge a\). Present a vertex \(x\) adjacent to every vertex of \(I_1\). Every $t$-fold online coloring algorithm needs to use a set of $t$ colors on $x$ that is disjoint from $\col(I_1)$.

Apply Lemma~\ref{lem:union} to find a choice of the side  call it $I$, which will be either the side contains $I_1 \cup I_2$ or the side contains $I_2 \cup \{x\}$. Lemma~\ref{lem:union} implies $|\col(I)| \ge a+s$ as desired. To verify the number of vertices, if \(m_j\) bounds the number of vertices, then \(m_0=1\) and \(m_j\le2m_{j-1}+1\), giving \(m_j\le2^{j+1}-1\). 
\end{proof}

Sine $K$ is our colors goal, a construction may stop immediately when \(K\) distinct online colors have appeared.   A color is \emph{two-sided} in a bipartite graph with designated sides \(A,B\) if it belongs to \(\col(A)\cap\col(B)\).

\begin{lemma}\label{lem:two-sided}
Let \(K\ge2\) and \(1\le t\le q\le K\). There exists a constant $C>0$ such that  for every  \(t\)-fold coloring algorithm, there is an adaptive bipartite graph construction with at most
    $CK\,2^{Cq/t}$ 
vertices which forces the algorithm to either  use \(K\) online colors globally or use at least \(q\) two-sided colors.
\end{lemma}

\begin{proof}
Use Lemma~\ref{lem:bipartite-independent} with
 $j=\ceil{\frac{2q-t}{\ceil{t/2}}}=O(q/t)$.
Present \(K\) disjoint copies of that construction, call them $G_1, .., G_K$, each using
\(2^{O(q/t)}\) vertices,  where the two sides of $G_i$ are denoted by $(A_i, B_i)$. Without loss of generality, assume the designated side is $A_i$, hence $\col(A_i) \ge 2q$ for every $1 \le i \le K$. Throughout the proof suppose the global number of colors remains \(p\le K-1\); otherwise we stop successfully. 

Form a bipartite incidence graph. Its left side consists of the
\(K\) constructed graphs, and its right side consists of the \(p\)
online colors used globally. Join a graph to a color if that color
occurs in its designated side. Every left degree is at
least \(2q\). Apply Lemma~\ref{lem:degree-count} with \(a=K>p=b\),
\(d=2q\), and \(i=2\), gives at least \(2q\) distinct colors each
occurring in the designated sides of at least two different graphs.

We now apply a probabilistic argument to show that one can re-orient the two sides of each $G_i$ so that the overall graph contains $q$ two-sided colors under the new orientation. For each $A_i$ we flip an independent coin to decide it will be on the left or right side. Since each of the $2q$ colors appears in at least two distinct $A_i$'s from two disjoint bipartite graph $G_i$'s, each color has a probability of at least $1/2$ to be two-sided. Therefore, in expectation there will be $q$ two-sided colors, and this implies the existence of such a desired orientation. 
\end{proof}

\subsection{Simulation of \(q\)-fold coloring by \(t\)-fold coloring}
\label{sec:simulation}

For a vertex \(v\) in an online graph with arrival order \(\pi\), define its
\emph{post-degree} as the number of its future neighbors. We say a construction has size bound \(N\) and post-degree bound \(D\) if it uses at most $N$ vertices and every vertex has post-degree at most $D$.

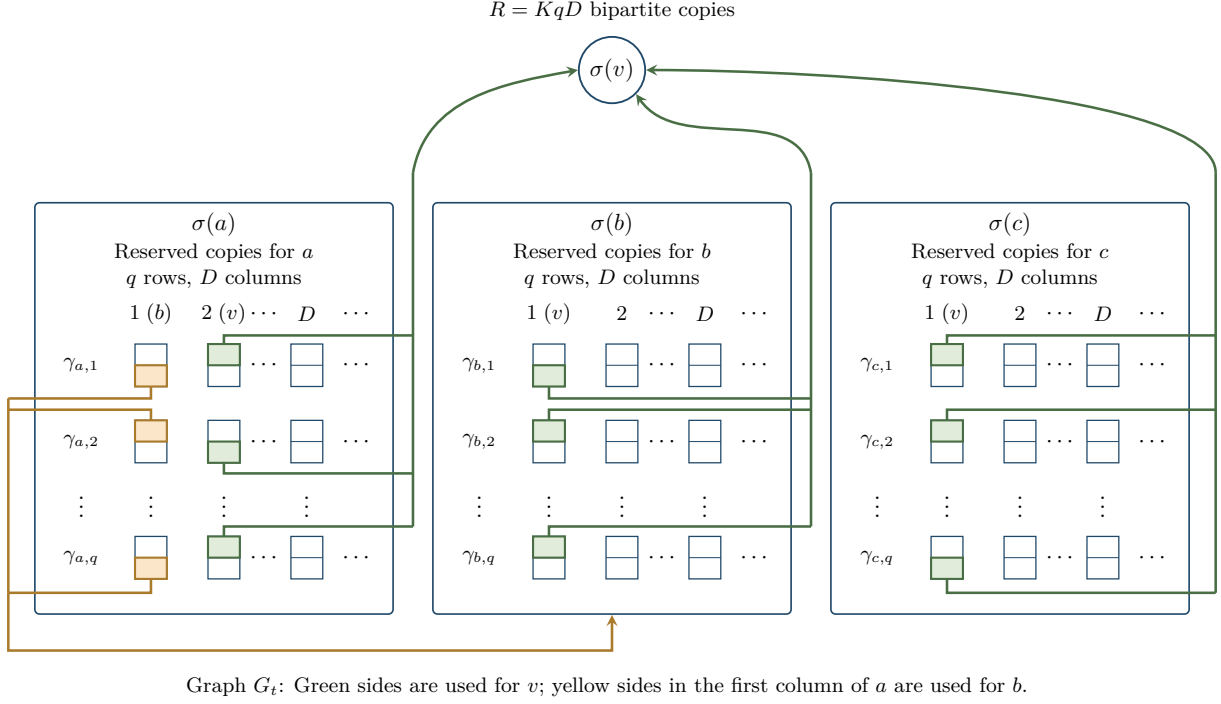
\begin{figure}[htbp!]
\centering
\begingroup
\definecolor{simnavy}{RGB}{31,72,105}
\definecolor{simgreen}{RGB}{225,237,218}
\definecolor{simyellow}{RGB}{251,230,201}
\definecolor{simgreenedge}{RGB}{73,110,68}
\definecolor{simyellowedge}{RGB}{169,123,43}
\resizebox{\linewidth}{!}{%
\begin{tikzpicture}[x=1cm,y=1cm,font=\small,
  collection/.style={draw=simnavy,line width=.7pt,
    rounded corners=2pt,fill=white,inner sep=4pt},
  representative/.style={circle,draw=simnavy,line width=.8pt,
    fill=white,minimum size=1cm,inner sep=2pt},
  greenbranch/.style={draw=simgreenedge,line width=1.1pt},
  yellowbranch/.style={draw=simyellowedge,line width=1.1pt},
  greenjoin/.style={draw=simgreenedge,line width=1.1pt,->,>=stealth},
  yellowjoin/.style={draw=simyellowedge,line width=1.1pt,->,>=stealth}]

  \node[representative] (sigv) at (0,5.1) {$\sigma(v)$};
  \node[font=\footnotesize] at (0,6)
    {$R=KqD$ bipartite copies};

  \foreach \w/\xx/\first/\second in {
    a/-6/{1\;(b)}/{2\;(v)},
    b/0/{1\;(v)}/{2},
    c/6/{1\;(v)}/{2}} {
    \begin{scope}[shift={(\xx,0)}]
      \node[collection,minimum width=5.4cm,minimum height=6.2cm]
        (box-\w) at (0,0) {};
      \node[font=\small\bfseries] at (0,2.77) {$\sigma(\w)$};
      \node[font=\footnotesize] at (0,2.34)
        {Reserved copies for $\w$};
      \node[font=\footnotesize] at (0,1.96)
        {$q$ rows, $D$ columns};
      \node[font=\footnotesize] at (-.95,1.43) {$\first$};
      \node[font=\footnotesize] at (.15,1.43) {$\second$};
      \node[font=\footnotesize] at (1.4,1.43) {$D$};
      \node[inner sep=0pt] at (.78,1.43) {$\cdots$};
      \node[inner sep=0pt] at (2.17,1.43) {$\cdots$};

      \foreach \i/\yy in {1/.65,2/-.5,q/-2.25} {
        \node[font=\footnotesize] at (-2,\yy)
          {$\gamma_{\w,\i}$};
        \foreach \cx in {-.95,.15,1.4} {
          \draw[draw=simnavy,line width=.55pt,fill=white]
            (\cx-.24,\yy-.32) rectangle (\cx+.24,\yy+.32);
          \draw[draw=simnavy,line width=.4pt]
            (\cx-.24,\yy)--(\cx+.24,\yy);
        }
        \node[inner sep=0pt] at (.78,\yy) {$\cdots$};
        \node[inner sep=0pt] at (2.17,\yy) {$\cdots$};
      }
      \foreach \cx in {-2,-.95,.15,1.4}
        \node[inner sep=0pt] at (\cx,-1.375)
          {$\vcenter{\hbox{$\vdots$}}$};
    \end{scope}
  }

  \foreach \xx/\yy/\lo/\hi in {
    -5.85/.65/0/.32, -5.85/-.5/-.32/0, -5.85/-2.25/0/.32,
    -.95/.65/-.32/0, -.95/-.5/0/.32, -.95/-2.25/0/.32,
    5.05/.65/0/.32, 5.05/-.5/0/.32, 5.05/-2.25/-.32/0} {
    \path[fill=simgreen,draw=simgreenedge,line width=.9pt]
      (\xx-.24,\yy+\lo) rectangle (\xx+.24,\yy+\hi);
  }
  \foreach \yy/\lo/\hi in {
    .65/-.32/0, -.5/0/.32, -2.25/-.32/0} {
    \path[fill=simyellow,draw=simyellowedge,line width=.9pt]
      (-6.95-.24,\yy+\lo) rectangle (-6.95+.24,\yy+\hi);
  }

  \foreach \start/\lane in {.97/1.10,-.82/-.98,-1.93/-1.78}
    \draw[greenbranch] (-5.85,\start)--(-5.85,\lane)--(-3,\lane);
  \foreach \start/\lane in {.33/.15,-.18/-.02,-1.93/-1.78}
    \draw[greenbranch] (-.95,\start)--(-.95,\lane)--(3,\lane);
  \foreach \start/\lane in {.97/1.10,-.18/-.02,-2.57/-2.78}
    \draw[greenbranch] (5.05,\start)--(5.05,\lane)--(9.1,\lane);

  \draw[greenbranch] (-3,-1.78)--(-3,3.55);
  \draw[greenbranch] (3,-1.78)--(3,3.55);
  \draw[greenbranch] (9.1,-2.78)--(9.1,3.55);
  \draw[greenjoin] (-3,3.55) to[out=80,in=190] (sigv.west);
  \draw[greenjoin] (3,3.55) to[out=100,in=305] (sigv.south east);
  \draw[greenjoin] (9.1,3.55)
    .. controls (9.1,4.75) and (3.2,5.1) .. (sigv.east);

  \foreach \start/\lane in {.33/.15,-.18/-.02,-2.57/-2.78}
    \draw[yellowbranch] (-6.95,\start)--(-6.95,\lane)--(-9.1,\lane);
  \draw[yellowbranch] (-9.1,.15)--(-9.1,-2.78);
  \draw[yellowjoin] (-9.1,-2.78)--(-9.1,-3.65)
    --(0,-3.65)--(box-b.south);
  \node[font=\footnotesize,align=center] at (0,-4.2) {
    Graph $G_t$: Green sides are used for $v$; yellow sides in the first column of $a$
    are used for $b$. };
\end{tikzpicture}%
}
\endgroup
\caption{The online simulation. Each $\sigma(w)$ consists of
$R=KqD$ bipartite copies. The displayed array consists of $qD$
reserved copies arranged in $q$ rows and $D$ columns: each column is reserved for simulating a future vertex in $G_q$ (the notation $(x)$ in the figure indicates the corresponding column is reserved for a future vertex $x \in V(G_q)$ that is to be simulated in $G_t$), and every bipartite graph in row $i$ contains the two-sided color $\gamma_{w,i}$. The online simulation then has the flexibility of choosing either side. }
\label{fig:simulation-edges}
\end{figure}

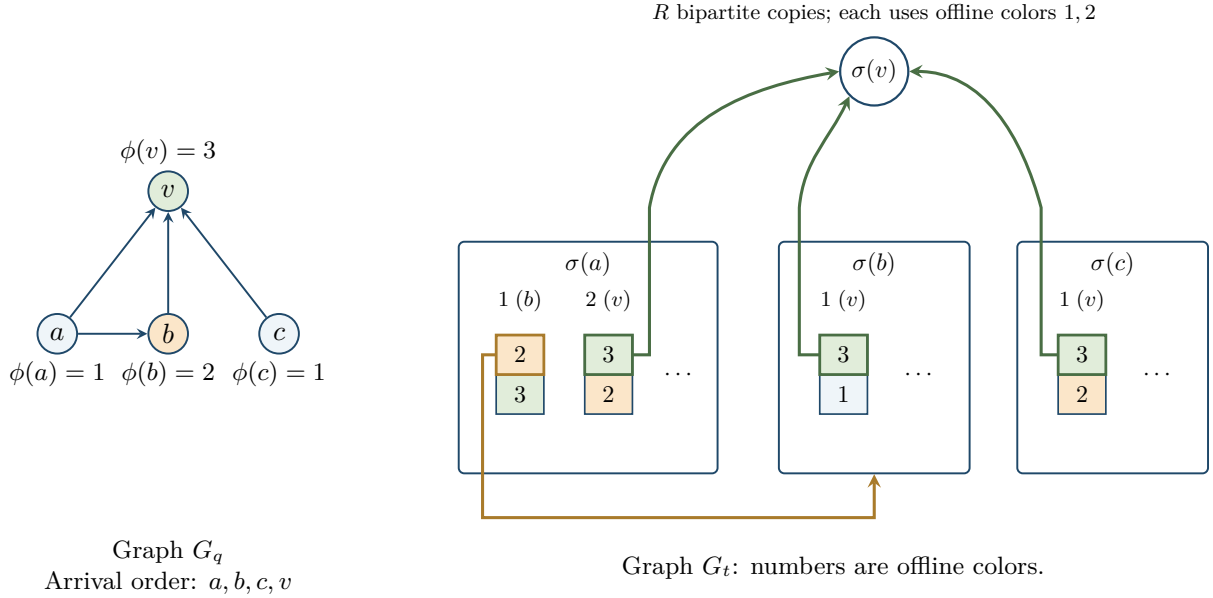
\begin{figure}[htbp!]
\centering
\begingroup
\definecolor{simnavy}{RGB}{31,72,105}
\definecolor{simlabel1}{RGB}{239,245,249}
\definecolor{simlabel2}{RGB}{251,230,201}
\definecolor{simlabel3}{RGB}{225,237,218}
\definecolor{simedge2}{RGB}{169,123,43}
\definecolor{simedge3}{RGB}{73,110,68}
\newcommand{\simcolumn}[4]{%
  \begin{scope}[shift={(#1,#2)}]
    \path[fill=simlabel#3] (-.35,0) rectangle (.35,.58);
    \path[fill=simlabel#4] (-.35,-.58) rectangle (.35,0);
    \draw[simnavy,line width=.6pt] (-.35,-.58) rectangle (.35,.58);
    \draw[simnavy,line width=.45pt] (-.35,0)--(.35,0);
    \draw[draw=simedge#3,line width=1pt]
      (-.35,0) rectangle (.35,.58);
    \node[font=\small,inner sep=0pt] at (0,.29) {$#3$};
    \node[font=\small,inner sep=0pt] at (0,-.29) {$#4$};
  \end{scope}%
}
\resizebox{\linewidth}{!}{%
\begin{tikzpicture}[x=1cm,y=1cm,font=\small,
  collection/.style={draw=simnavy,line width=.7pt,
    rounded corners=2pt,fill=white,inner sep=4pt},
  representative/.style={circle,draw=simnavy,line width=.8pt,
    fill=white,minimum size=1cm,inner sep=2pt},
  virtual/.style={circle,draw=simnavy,line width=.7pt,
    minimum size=.5cm,inner sep=1pt},
  greenjoin/.style={draw=simedge3,line width=1.1pt,->,>=stealth},
  yellowjoin/.style={draw=simedge2,line width=1.1pt,->,>=stealth}]

  \begin{scope}[yshift=-.5cm]
    \node[virtual,fill=simlabel1] (va) at (-7.45,1) {$a$};
    \node[virtual,fill=simlabel2] (vb) at (-6.05,1) {$b$};
    \node[virtual,fill=simlabel1] (vc) at (-4.65,1) {$c$};
    \node[virtual,fill=simlabel3] (vv) at (-6.05,2.8) {$v$};
    \draw[simnavy,line width=.7pt,->,>=stealth] (va)--(vb);
    \foreach \w in {va,vb,vc}
      \draw[simnavy,line width=.7pt,->,>=stealth] (\w)--(vv);
    \node[font=\footnotesize] at (-7.45,.5) {$\phi(a)=1$};
    \node[font=\footnotesize] at (-6.05,.5) {$\phi(b)=2$};
    \node[font=\footnotesize] at (-4.65,.5) {$\phi(c)=1$};
    \node[font=\footnotesize] at (-6.05,3.3) {$\phi(v)=3$};
  \end{scope}

  \begin{scope}[shift={(.8,.2)},scale=.86,transform shape]
    \node[representative] (sigv) at (2.4,4.2) {$\sigma(v)$};
    \node[font=\footnotesize,align=center] at (2.4,5.05)
      {$R$ bipartite copies; each uses offline colors $1,2$};

    \node[collection,minimum width=3.8cm,minimum height=3.4cm]
      (box-a) at (-1.8,0) {};
    \node[collection,minimum width=2.8cm,minimum height=3.4cm]
      (box-b) at (2.4,0) {};
    \node[collection,minimum width=2.8cm,minimum height=3.4cm]
      (box-c) at (5.9,0) {};
    \foreach \w/\xx in {a/-1.8,b/2.4,c/5.9}
      \node[font=\small\bfseries] at (\xx,1.38) {$\sigma(\w)$};
    \node[font=\footnotesize] at (-2.8,.83) {$1\;(b)$};
    \node[font=\footnotesize] at (-1.5,.83) {$2\;(v)$};
    \node[font=\footnotesize] at (1.95,.83) {$1\;(v)$};
    \node[font=\footnotesize] at (5.45,.83) {$1\;(v)$};

    \simcolumn{-2.8}{-.25}{2}{3}
    \simcolumn{-1.5}{-.25}{3}{2}
    \simcolumn{1.95}{-.25}{3}{1}
    \simcolumn{5.45}{-.25}{3}{2}
    \foreach \xx in {-.45,3.08,6.58}
      \node[inner sep=0pt] at (\xx,-.25) {$\cdots$};

    \draw[greenjoin] (-1.15,.04)--(-.9,.04)--(-.9,2.2)
      to[out=80,in=190] (sigv.west);
    \draw[greenjoin] (1.6,.04)--(1.3,.04)--(1.3,2.2)
      to[out=85,in=245] (sigv.south west);
    \draw[greenjoin] (5.1,.04)--(4.85,.04)--(4.85,2.2)
      to[out=105,in=0] (sigv.east);
    \draw[yellowjoin] (-3.15,.04)--(-3.35,.04)--(-3.35,-2.35)
      --(2.4,-2.35)--(box-b.south);
  \end{scope}

  \node[font=\footnotesize,align=center] at (-6.05,-2.45) {
    Graph $G_q$\\
    Arrival order: $a,b,c,v$};

  \node[font=\footnotesize,align=center] at (2.348,-2.45) {
    Graph $G_t$: numbers are offline colors.};
\end{tikzpicture}%
}
\endgroup
\caption{The corresponding offline three-coloring. Here each bipartite block corresponds to a column in Figure \ref{fig:simulation-edges}. The flexibility of choosing either side allows the graph $G_t$ to be three-colorable as long as $G_q$ is. If $\phi$ is a three-coloring on $G_q$, then the bipartite graph $\sigma(v) \subseteq G_t$ will be colored by the two colors in $\{1,2,3\} \backslash \{\phi(v)\}$.}
\label{fig:simulation-coloring}
\end{figure}

\begin{lemma}[Simulation lemma]\label{lem:simulation}
    Let \(1\le t<q\le K\), $D > 0$. There exists a constant $C>0$ such that, every three-colorable graph construction with size bound \(N\) and post-degree bound \(D\) forcing $q$-fold coloring to use $K$ colors can be simulated by a three-colorable graph construction forcing $t$-fold coloring to use $K$ colors, with
size bound \(N'\) and post-degree bound \(D'\) satisfying
\begin{equation}
 N'\le CK^2q\,ND\,2^{Cq/t},
 \qquad
 D'\le CK^2q\,D\,2^{Cq/t}.
 \label{eq:simulation-bounds}
\end{equation}
\end{lemma}

\begin{proof}
Let $G_q$ denote the given online \(q\)-fold construction, our goal is to build an online graph $G_t$ to simulate $G_q$. 
For each vertex $v \in V(G_q)$, let $\sigma(v)\subseteq V(G_t)$ denote the collection of vertices in $G_t$ that plays the role of simulating the single vertex $v \in V(G_q)$. Every $\sigma(v)$ consists of 
\begin{equation}
 R=KqD
 \label{eq:number-copies}
\end{equation}
copies of the \(t\)-fold bipartite construction from
Lemma~\ref{lem:two-sided}. By Lemma~\ref{lem:two-sided}, each copy has at most \(B=CK2^{Cq/t}\) vertices.
The factor \(D\) reserves a separate copy for each possible future
neighbor of a simulated vertex. This permits different copies to
be oriented differently in an offline three-coloring, while also
controlling the post-degree in the new construction, see Figure \ref{fig:simulation-edges} and Figure \ref{fig:simulation-coloring} for illustrations.

Throughout the simulation, we will maintain the following properties:
\begin{itemize}
    \item  X: for every $v\in G_q$,  $\col(\sigma(v)) \supseteq \col(v)$;

    \item Y: for each color $i \in \col(v)$, there are $D$ copies in $\sigma(v)$ each of which contains the color $i$ as a two-sided color. Furthermore, these $D$ copies for distinct $i$ are disjoint, see Figure \ref{fig:simulation-edges}.
\end{itemize}

To simulate a general online vertex $v \in V(G_q)$, we construct $\sigma(v)$ as described before and will describe the online edge connection below. If \(K\) colors appear globally in $G_t$, stop. Hence, we assume the number of colors used globally is at most $K-1$. 

\paragraph{Online edge connection.} Since the simulation can be done inductively, we assume properties hold for pre-neighbors of $v$. Let $w \in V(G_q)$  be such a neighbor. We look at $\sigma(w)$. By the degree constraint $D$ of graph $G_q$ and the property Y, there are $q$ \emph{unused} copies (correspond to a column in Figure \ref{fig:simulation-edges}) in $\sigma(w)$ each contains a two-sided color $i$ for $i\in \col(w)$. In each copy, we can choose a vertex of color $i$ from either side to connect to all vertices of $\sigma(v)$. This will make these copies become \emph{used} and will not be used later. This finishes the online edge connection. 

\paragraph{Properties X and Y for $\sigma(v)$.} The online edge connection clearly simulates the coloring constraints of $v$ on $\sigma(v)$. Hence, $\sigma(v)$ uses at least $q$ colors that are disjoint from all $\col(w)$ for $w$ as pre-neighbor of $v$. 

Apply Lemma \ref{lem:two-sided}, we get that each copy in $\sigma(v)$ contains $q$ two-sided colors. Apply Lemma \ref{lem:degree-count} similarly as in the proof for Lemma \ref{lem:two-sided} with parameters
\[
 a=R,\qquad b=p,\qquad d=q,\qquad i=qD,
\]
we get that at least $q$ colors each occurs in at least $qD$ copies in $\sigma(v)$. By renaming if necessary, we identify these $q$ colors as the set $\col(v)$. Then, property Y can be realized greedily: for color $i \in \col(v)$, successively choose $D$ copies from its $qD$ copies, disjoint from all previous choices.

\paragraph{Offline three-colorability of $G_t$.} Fix a proper three-coloring
$\phi:V(G_q)\to\{1,2,3\}$.
We color $G_t$ so that every bipartite copy in $\sigma(w)$
uses only the two colors different from $\phi(w)$;
see Figure~\ref{fig:simulation-coloring}.
Process the collections $\sigma(w)$ in reverse order of the
corresponding vertices of $G_q$. Consider a bipartite copy in $\sigma(w)$. If this copy was used for connections to a later collection $\sigma(v)$, give its selected side color $\phi(v)$ and its other side the unique color in $\{1,2,3\}\setminus\{\phi(w),\phi(v)\}$. This is well-defined because $wv\in E(G_q)$ implies $\phi(w)\ne\phi(v)$, and each copy is used for connections
to at most one later collection. If the copy has no such connection, color its two sides arbitrarily with the two colors different from $\phi(w)$. Clearly, this coloring is a legal three-coloring on $G_t$.

\paragraph{Size and post-degree.} 
$N'$ is clear: $N' \le N \cdot R \cdot B$. For $D'$, our construction makes sure that each vertex in $G_t$ is connected to at most one $\sigma(v)$, besides, it may be connected to at most $B-1$ vertices in its bipartite copy. Hence, the post-degree $D'$ is at most $R\cdot B +B$. By choosing the constant $C>0$ suitably, both claimed upper bounds hold. 
\end{proof}

\subsection{Finishing the proof}\label{sec:lower-final}

\begin{proof}[Proof of Theorem~\ref{thm:deterministic}]
Fix a sufficiently large integer $K$. We first construct a
three-colorable graph that forces $K$ colors in the ordinary online
coloring model, while bounding its number of vertices.

Set
\[
 \ell=\log K,\qquad
 \beta=\lceil\ell\rceil,\qquad
 h=\lceil\log_\beta K\rceil,
 \qquad
 t_h = K, \ 
 t_i=\beta^i \quad(0\le i\le h-1).
\]
These are integers satisfying
\[
 t_0=1,\qquad t_h=K,\qquad
 t_i<t_{i+1},\qquad
 \frac{t_{i+1}}{t_i}\le\beta
 \quad(0\le i<h).
\]
Furthermore,
\[
 h\le\frac{\ell}{\log\ell}+1
   =O\!\left(\frac{\ell}{\log\ell}\right).
\]

In the $K$-fold coloring model, one vertex forces $K$ colors.
This construction has size bound $N_h=1$ and post-degree bound
$D_h=1$, where the latter is an upper bound on its actual
post-degree zero.

Apply Lemma~\ref{lem:simulation} successively with
\[
 q=t_{i+1},\qquad t=t_i,
 \qquad i=h-1,h-2,\ldots,0.
\]
The target number of colors remains the same integer $K$
throughout these applications.
Let $N_i,D_i$ denote the size and post-degree bounds of the
resulting $t_i$-fold construction. Every resulting construction is three-colorable and forces $K$ colors.

Choose an absolute constant $C_0\ge1$ sufficiently large and put
\[
 F=C_0K^3\,2^{C_0\beta}.
\]
Since $q\le K$ and $q/t\le\beta$ in every application,
Lemma~\ref{lem:simulation} gives
\begin{equation}
 D_i\le F D_{i+1},
 \qquad
 N_i\le F N_{i+1}D_{i+1}.
 \label{eq:lower-recurrence}
\end{equation}
By induction with the base case $D_h=1$ and $N_h=1$, we obtain that for $s=h-i$,
\begin{equation}
 D_i\le F^s,
 \qquad
 N_i\le F^{s(s+1)/2}.
 \label{eq:lower-induction}
\end{equation}

Because $\beta=\lceil\ell\rceil$, we have
 $\log F=\log C_0+3\ell+C_0\beta=O(\ell)$.
Consequently,
\[
 \log N_0
 \le \frac{h(h+1)}2\log F
 =O\!\left(\frac{\ell^3}{(\log\ell)^2}\right).
\]
Since $t_0=1$, this is an ordinary online coloring construction.
Thus there is an absolute constant $A\ge1$ such that, for every
sufficiently large $K$, it forces $K$ colors using at most
\begin{equation}
 N_0\le
 2^{\,A(\log K)^3/(\log\log K)^2}
 \label{eq:size-K}
\end{equation}
vertices.

We now express the bound in terms of the number of vertices.
Let
\[
 x=\log n,\qquad
 s=x^{1/3}(\log x)^{2/3},
\]
and choose an absolute constant $c>0$ sufficiently small that
 $128Ac^3\le1$.
Set
 $K=\left\lceil 2^{cs}\right\rceil$ and 
 $\ell=\log K$.
For all sufficiently large $n$,
\[
 cs\le\ell\le cs+1\le2cs,
 \qquad
 \log\ell\ge\frac14\log x.
\]
Applying~\eqref{eq:size-K}, we obtain
\[
 \log N_0
 \le A\frac{\ell^3}{(\log\ell)^2}
 \le
 A\frac{(2cs)^3}{(\tfrac14\log x)^2}
 =128Ac^3x
 \le x.
\]
Therefore $N_0\le n$, while the construction forces at least
 $K\ge
 2^{c(\log n)^{1/3}(\log\log n)^{2/3}}$
colors.
\end{proof}

\section{A randomized upper bound}\label{sec:upper}

By Remark \ref{rmk:after_thm}, we assume that $n$, the number of vertices of an online graph $G$, is known.

\subsection{An attempted algorithm and an overview of the design}\label{sec:attempt}

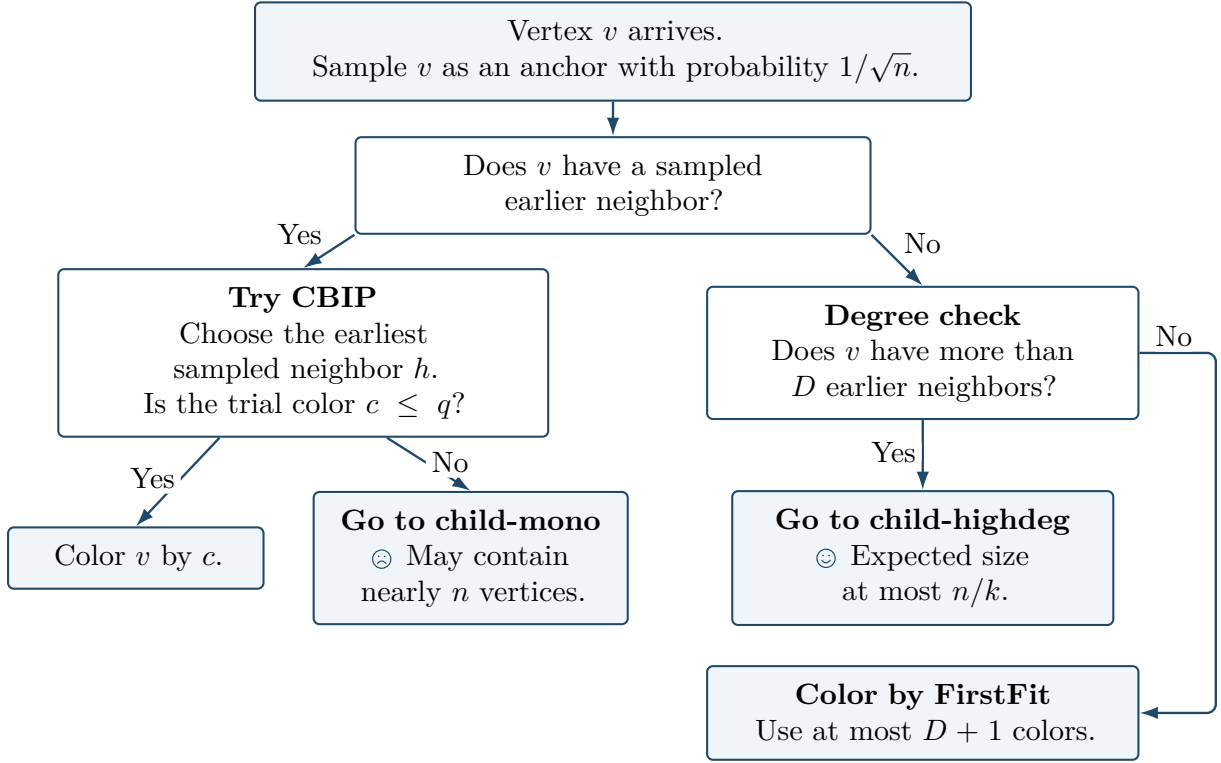
\begin{figure}[htbp!]
\centering
\begingroup
\definecolor{attemptblue}{RGB}{31,72,105}
\definecolor{attemptpale}{RGB}{239,245,249}
\newcommand{\attemptface}[1]{%
  \tikz[baseline=-.5ex,x=1ex,y=1ex]{%
    \draw[attemptblue,line width=.4pt] (0,0) circle (.72);
    \fill[attemptblue] (-.24,.22) circle (.065);
    \fill[attemptblue] (.24,.22) circle (.065);
    \ifnum#1=1\relax
      \draw[attemptblue,line width=.4pt]
        (-.35,-.12) .. controls (-.15,-.43) and (.15,-.43) ..
        (.35,-.12);
    \else
      \draw[attemptblue,line width=.4pt]
        (-.35,-.38) .. controls (-.15,-.07) and (.15,-.07) ..
        (.35,-.38);
    \fi
  }%
}
\hyphenpenalty=10000
\exhyphenpenalty=10000
\resizebox{\linewidth}{!}{%
\begin{tikzpicture}[x=1cm,y=1cm,font=\small,
  abox/.style={
    draw=attemptblue,line width=.7pt,
    rounded corners=2pt,fill=white,align=center,
    inner xsep=5pt,inner ysep=6pt},
  aend/.style={abox,fill=attemptpale},
  aarrow/.style={
    -{Latex[length=2.2mm,width=1.5mm]},
    draw=attemptblue,line width=.8pt,
    rounded corners=3pt},
  alabel/.style={
    font=\small,fill=white,inner sep=2pt}]

  \node[aend,text width=8.2cm] (arrive) at (0,0) {
    Vertex $v$ arrives.\\
    Sample $v$ as an anchor with probability $1/\sqrt n$.
  };

  \node[abox,text width=5.8cm] (anchor) at (0,-1.6) {
    Does $v$ have a sampled\\earlier neighbor?
  };

  \node[abox,text width=5.5cm] (cbip) at (-3.7,-3.6) {
    \textbf{Try CBIP}\\
    Choose the earliest sampled neighbor $h$.\\
    Is the trial color $c\le q$?
  };

  \node[abox,text width=4.8cm] (degree) at (3.7,-3.6) {
    \textbf{Degree check}\\
    Does $v$ have more than $D$ earlier neighbors?
  };

  \node[aend,text width=2.7cm] (accept) at (-5.7,-6.05) {
    Color $v$ by $c$.
  };

  \node[aend,text width=3.4cm] (mono) at (-1.7,-6.05) {
    \textbf{Go to child-mono}\\
    \attemptface{0}\enspace May contain nearly $n$ vertices.
  };

  \node[aend,text width=4.2cm] (high) at (3.7,-6.05) {
    \textbf{Go to child-highdeg}\\
    \attemptface{1}\enspace Expected size at most $n/k$.
  };

  \node[aend,text width=4.8cm] (ff) at (3.7,-7.9) {
    \textbf{Color by FirstFit}\\
    Use at most $D+1$ colors.
  };

  \draw[aarrow] (arrive)--(anchor);

  \draw[aarrow] (anchor.south west)--
    node[alabel,pos=.5,above left] {Yes}
    (cbip.north);

  \draw[aarrow] (anchor.south east)--
    node[alabel,pos=.5,above right] {No}
    (degree.north);

  \draw[aarrow] ($(cbip.south)+(-1,0)$)--
    node[alabel,pos=.45,left] {Yes}
    (accept.north);

  \draw[aarrow] ($(cbip.south)+(1,0)$)--
    node[alabel,pos=.45,right] {No}
    (mono.north);

  \draw[aarrow] (degree.south)--
    node[alabel,pos=.45,left] {Yes}
    (high.north);

  \draw[aarrow] (degree.east)--
    node[alabel,pos=.45,above] {No}
    (7.2,-3.6) |- (ff.east);

\end{tikzpicture}%
}
\endgroup
\caption{An attempted algorithm on $n$ vertices, with
$D=k\sqrt n$. The obstacle is that child-mono may contain
too many vertices.}
\label{fig:attempted-algorithm}
\end{figure}

Consider the algorithm described in Figure \ref{fig:attempted-algorithm}. One can see that it is a similar but slightly different version of Vishwanathan's randomized algorithm discussed in Figure \ref{fig:vishwanathan-upper}.

The idea is the following:
\begin{itemize}
    \item sample roughly $\sqrt{n}$ anchors by independently sampling each vertex to be an anchor with probability roughly $1/\sqrt{n}$;
    
    \item for anchor's neighborhood, color it using CBIP but with only $q$ colors budget, and for vertices exceeding that budget, send them to a recursive processing that we call child-mono;

    \item for vertices without any anchor: if it has degree at most $D$, we color it by FirstFit using at most $D+1$ colors, and for high-degree vertices, we send them again to a recursive processing that we call child-highdeg. Setting $D = k\cdot \sqrt{n}$ would guarantee that high-degree vertices are unlikely to have no sampled-anchor, hence the child-highdeg receives only about $n/k$ vertices in expectation. 
\end{itemize}

Choosing $q,k=\Theta(1)$ guarantees we use only $O(\sqrt{n})$ colors locally for both CBIP and FirstFit, and our goal would be achieved if we can show the children also use $O(\sqrt{n})$ colors: for this we would like to have a compression effect, that is, the size of each child is only a fraction of the whole instance. This can already be made true for child-highdeg, but there is no such obvious guarantee for child-mono.

\paragraph{Compressing the input to child-mono.}
We will show that although vertices go into child-mono might be many, nonetheless, they can be divided into groups, where each group is a subset of some offline-monochromatic class of $G$, and the total number of such offline-monochromatic classes is only a fraction of $n$. This allows a compression effect similar to Blum's work as discussed before. We can think of each of those groups in child-mono as a single \emph{virtual} vertex, in that sense we can show that the size of the virtual graph for child-mono is only a fraction of the size of $G$. Each group (virtual vertex) can be colored using a single color in offline coloring, while in online coloring each group may still require many colors. 

\paragraph{Vertex-label pair $(v,\ell)$.}
Following the discussion, then, each vertex in child-mono should be viewed as a pair $(v,\ell)$ where $\ell$ denotes the group (i.e., a subset of some offline-monochromatic class) it belongs to. In other words, child-mono is a graph on virtual vertices $\ell$'s. Let us temporarily denote this online virtual graph by $M$. Once we take this view of a vertex in child-mono, because our algorithm is intended to be recursive, we need to view every vertex going into each recursive instance this way, and every child will be viewed as a virtual graph on labels. Though, when we do recursive online coloring, we still need to color the real vertex $v$, not label $\ell$.

For example, in the original instance $G$, each vertex can be viewed as $(v,v)$ where $v$ is its own label. When a vertex $v$ goes into the child-mono $M$ of $G$, it will appear as $(v,\ell)$ where $\ell$ denotes a label for some offline-monochromatic class of $G$. If the vertex is further sent to the child-mono of $M$, denoted by $\hat{M}$, then $v$ would appear as $(v,\hat{\ell})$ where $\hat{\ell}$ denotes a label for some offline-monochromatic class of \emph{labels} in the virtual graph $M$, and hence $\hat{M}$ would be a virtual graph on those $\hat{\ell}$'s, etc.

\paragraph{How does this affect anchor-sampling?}
Once we take the virtual graph view, then in every recursive instance, it makes sense to do the anchor-sampling with respect to the virtual vertex (i.e., labels), and each label is only sampled once: this will guarantee we have the desired expected number of anchors.

\paragraph{How does this affect degree-control?} 
Similarly, we test the degree of a vertex $v$ on the virtual graph instead of its real degree in $G$. Since we do anchor-sampling with respect to the virtual vertex (i.e., label), the high-degree part does not change: it is still unlikely for a high virtual-degree vertex to have no anchored neighbor. However, for low-virtual-degree vertices, it is different. 

Consider $(v,\ell)$ to have virtual-degree at most $D$ in a virtual graph, that is, $\ell$ is connected to at most $D$ distinct other labels, say $\ell^{(i)}$ for $i=1,\ldots, t$ where $t\le D$. Now, if we try FirstFit for $v$, we would like $v$ to be colored by a color that is absent from  $\cup_{i=1}^t \col(\ell^{(i)})$. But since each $\ell^{(i)}$ is a group of vertices, which means each $\col(\ell^{(i)})$ is a set of colors instead of a single color, it is no longer necessarily true that $v$ can still be colored by at most $D+1$ colors greedily. Therefore, in order to have certain control on the colors used on every label $\ell$, we choose a separate palette for each $\ell$ (of constant size, say $L$), and then we do the following: if $(v,\ell)$ can be colored via FirstFit using the palette for $\ell$, we color it, otherwise we sent it to another recursive stream that we call child-lowdeg. Now, in order to make sure child-lowdeg is small, we choose the size-$L$ palette for each $\ell$ \emph{randomly} and independently from a large shared palette of size $K\approx DL$. Since $(v,\ell)$ has virtual degree at most $D$, we can then show that with high probability $(v,\ell)$ can be colored using the palette of $\ell$, in other words, the virtual graph of child-lowdeg would be small, as desired.

\subsection{Virtual graphs and coloring rules}\label{sec:definitions}

We now write $H$ for the current virtual graph and $M$ for an
upper bound on its number of labels.
The input consists of pairs $(v,\ell)$.
Let $V_\ell$ be the group of actual vertices carrying label $\ell$.
We require each $V_\ell$ to be independent and the final virtual
graph to be three-colorable.
At the root, every vertex has its own label.

The vertices of $H$ are the labels received by the instance.
Two labels are adjacent when an edge between their actual vertices
has been revealed.
The virtual graph records all received pairs and revealed edges, even when a
vertex is colored by another palette or forwarded to a child.
Thus $H$ only grows, and later arrivals may add edges between
existing labels.
All degrees below are degrees in $H$.
Labels are ordered by their first appearance.

Maintain offline-monochromatic classes by identifying the common
neighbors of every edge of $H$ and taking the transitive closure.
These classes only merge.
The validity of these certificates is proved in
Lemma~\ref{lem:neighborhood-mono}.

\paragraph{Color sets and FirstFit.}
As in the lower-bound section, $\col(S)$ denotes the colors
assigned to vertices in $S$.
For an ordered shared palette $P$, write
\[
 \col_P(\ell)=\col(V_\ell)\cap P.
\]
Each group has a fixed color-table $T_\ell\subseteq P$ of size $L$.
FirstFit chooses the least color in
\begin{equation}
 T_\ell\setminus
 \bigcup_{h\in N_H(\ell)}\col_P(h).
 \label{eq:ff}
\end{equation}
It fails if this set is empty.
Only this rule uses colors from $P$, so
\begin{equation}
 \col_P(\ell)\subseteq T_\ell,\qquad
 |\col_P(\ell)|\le L.
 \label{eq:history}
\end{equation}

\paragraph{CBIP.}
A sampled label $h$ is an anchor, with its own palette of $q$ colors.
Use local color names $1,\ldots,q$ in each anchor palette;
the same name in different palettes represents different actual colors.
Let $\col_h(\ell)$ record the local colors accepted at $\ell$
from $h$'s palette.

For a request at $\ell\in N_H(h)$, consider its component in
\[
 J_h=H[N_H(h)],
\]
and let $Y$ be the side opposite $\ell$.
The CBIP trial color is
\begin{equation}
 c=\min\left(\mathbb N_{>0}\setminus
                  \bigcup_{u\in Y}\col_h(u)\right).
 \label{eq:cbip}
\end{equation}
Accept it if $c\le q$ and record it in $\col_h(\ell)$.
This is the usual CBIP rule with retained color sets
\cite{LST1989,BianchiEtAl2014,LiNarayanPankratov2022}.
All accepted colors are retained throughout the run;
a failed trial changes no color set.

The shared palette, the anchor palettes, and the palettes of
different recursive instances and runs are mutually disjoint.

\subsection{The full algorithm}\label{sec:algorithm}

Fix
\begin{equation}
 k=2^{16},\qquad L=8,\qquad q=2\ceil{\log k}+4=36.
 \label{eq:constants}
\end{equation}
For a label budget $M\ge1$, set
\begin{equation}
 D=k\ceil{\sqrt M},\qquad K=4DL.
 \label{eq:parameters}
\end{equation}
All logarithms are to base $2$.

Let $B_t(M)$ be one run with at most $M$ distinct labels and
remaining recursion depth $t$.
At $t=0$, give each group one private color.
For $t\ge1$, create the shared palette $P$ of $K$ colors,
start with empty virtual graph and color sets, and create three
child streams: child-mono, child-highdeg, and child-lowdeg.

At the first appearance of $\ell$, independently sample it as an
anchor with probability $1/\ceil{\sqrt M}$ and, if sampled,
reserve its $q$ anchor colors.
Independently sample
\[
 T_\ell\ \text{uniformly from }\binom{P}{L}.
\]
All choices are independent across labels and runs.
Later requests with the same label reuse its choices.

For forwarding to child-mono, each maintained class may have
a class label.
Create one only when needed and none exists.
On a class merger, retain the earliest existing class label for
future forwarding.
Previously forwarded vertices keep their child labels.

\begin{center}
\fbox{\begin{minipage}{\dimexpr\textwidth-2\fboxsep-2\fboxrule\relax}
\small
\textbf{Processing $(v,\ell)$ in $B_t(M)$.}

If $t=0$, give $v$ the private color of $\ell$ and stop.
Otherwise update $H$, its classes, and inherited class labels.
If $\ell$ is new, perform its anchor and color-table sampling.
Apply the following rules in order:
\begin{enumerate}[leftmargin=*,itemsep=5pt]
 \item \textbf{Try CBIP.}
 If $\ell$ has a sampled neighbor, choose the earliest one $h$
 and compute its CBIP trial color.
 On success, color $v$ from $h$'s palette and record the color.
 On failure, forward $v$ to child-mono under its class label,
 creating that label if necessary.

 \item \textbf{Degree check.}
 If there is no sampled neighbor and $\deg_H(\ell)>D$,
 forward $(v,\ell)$ to child-highdeg.

 \item \textbf{Try FirstFit with $T_\ell$.}
 Otherwise, $\deg_H(\ell)\le D$.
 If~\eqref{eq:ff} supplies a color, assign it to $v$ and
 record it in $\col_P(\ell)$.
 If not, forward $(v,\ell)$ to child-lowdeg.
\end{enumerate}
Stop after the applicable rule.
Every forwarded vertex is colored recursively before the next
actual vertex arrives.
\end{minipage}}
\end{center}

Each child receives the induced stream of actual vertices
forwarded to it, with their specified labels.
It uses the procedure $A_{t-1}$ below, with fresh independent
randomness.
The parent never uses the child's output in its decisions.

\paragraph{Unknown child sizes.}
The procedure $A_t$ starts with $B_t(1)$.
It tracks the cumulative number of distinct labels received.
Before a request makes this count exceed the current budget $M$,
double $M$ and start a fresh $B_t(M)$.
The new run has empty state, new palettes and choices, and fresh
child streams.
It considers only vertices arriving during that run and edges
between them.
Earlier vertices retain their colors.
Each run receives at most its budget's number of labels;
disjoint palettes handle edges across runs.
Empty streams are not initialized.

\paragraph{Root.}
Set $T=\ceil{\log(n+1)}$ and submit the $i$-th vertex as
$(v_i,i)$ to $B_T(n)$.

Fix an ordered input to one $B_t(M)$ run independently of its
random choices.
Write $m_{\mono},m_{\highdeg},m_{\lowdeg}$ for the numbers of
distinct labels received by its children. We proceed to bound these.

\subsection{Offline-monochromatic classes and child-mono}
\label{sec:mono}

In this subsection we use properties of CBIP and the three-colorability of the overall graph to control the size of child-mono.

\begin{lemma}\label{lem:neighborhood-mono}
Let $G$ be three-colorable and $v\in V(G)$.
Each side of every connected component of $G[N_G(v)]$
is an offline-monochromatic class.
\end{lemma}

\begin{proof}
Consider a length-two path $x,u,y$ in $G[N_G(v)]$.
The vertices $x,y$ are common neighbors of the edge $vu$,
so they receive the same color in every proper three-coloring, i.e., they belong to an offline-monochromatic class.
Since any two vertices on the same side of a connected bipartite
component are joined by an even path, the claim follows.
\end{proof}

\begin{lemma}[CBIP growth]\label{lem:growth}
On a bipartite graph, if CBIP uses color $a\ge2$ on a connected component, then each side of that component has at least
 $2^{\floor{a/2}-1}$ 
vertices.
\end{lemma}
\begin{proof}
We induct on $a$.
For $a=2,3$, the component contains an edge.
For $a\ge4$, put $j=a-2$.
At the occurrence of $a$, the opposite side contains an occurrence
of $j+1$ at some vertex $x$ and an occurrence of $j$ at some vertex
$y$.
When $j+1$ was accepted at $x$, there was an occurrence of $j$
at a vertex $z$ opposite $x$.
Thus $y,z$ lie on opposite sides of the current component.

At the later of these two specified occurrences of $j$,
the components of $y,z$ were distinct:
otherwise the earlier occurrence would have forbidden the later one.
Each component then had at least $2^{\floor{j/2}-1}$ vertices
on each side by induction.
Their later merger gives at least
 $2\cdot2^{\floor{j/2}-1}=2^{\floor{a/2}-1}$
vertices on each side.
\end{proof}

These two lemmas implies the following.

\begin{corollary}[CBIP failure]\label{cor:overflow}
    For the fixed parameter $q$, if in our algorithm CBIP needs to use  color $c>q$ on a vertex $v$, then the maintained offline-monochromatic class where $v$ belongs to contains at least $k$ vertices.
\end{corollary}

\begin{proof}
By Lemma~\ref{lem:neighborhood-mono} and  Lemma~\ref{lem:growth}, since $c\ge q+1=2\ceil{\log k}+5$, the side containing $v$
has at least
 $2^{\floor{c/2}-1}\ge2^{\ceil{\log k}+1}\ge k$
vertices, and they all lie in one
maintained offline-monochromatic class.
\end{proof}

These statements apply directly to $H$, whose vertices are labels.
In particular, every class label created for child-mono represents
a subset of a parent offline-monochromatic class containing at
least $k$ labels.

\begin{lemma}\label{lem:mono}
    $m_{\mono} \le M/k$,  and its virtual graph is
three-colorable.
\end{lemma}
\begin{proof}
By Corollary \ref{cor:overflow}, at each class-label creation, choose $k$ parent labels in its class.
The chosen sets are disjoint:
a later class containing any previously chosen label inherits
a class label and cannot cause another creation.
Since the parent has at most $M$ labels, there are at most
$M/k$ creations.

The three-colorability of virtual graph follows from the fact that each label denotes a subset of some offline-monochromatic class, and by definition each such class receives the same color in every fixed three-coloring, that is, the virtual vertex can be colored using the color for that class.
\end{proof}

\subsection{The high-degree and low-degree children}\label{sec:highlow}

\begin{lemma}\label{lem:high}
$\E m_{\highdeg}\le M/k$.
\end{lemma}
\begin{proof}
For a label $\ell$ that ever has degree greater than $D$,
let $S$ be its neighbor set at the first such request.
The input fixes $S$, and $|S|>D$.
If $\ell$ ever enters child-highdeg, none of $S$ can be sampled,
because its neighbor sets only grow.
Therefore
\[
 \Prb(\ell\text{ enters child-highdeg})
 \le
 \left(1-\frac1{\ceil{\sqrt M}}\right)^{|S|}
 \le e^{-|S|/\ceil{\sqrt M}}
 <e^{-k}\le\frac1k.
\]
Summing over at most $M$ labels proves the claim.
\end{proof}

\begin{lemma}\label{lem:low}
$\E m_{\lowdeg}\le M/k$.
\end{lemma}
\begin{proof}
Condition on all anchor choices.
The positions reaching FirstFit are now fixed independently of
the color-tables: they have no sampled neighbor and degree at
most $D$.

For a label $\ell$ reaching FirstFit, let $S_\ell$ be its neighbor
set at the last such position.
Then $|S_\ell|\le D$, and $S_\ell$ contains the neighbor sets
at all earlier FirstFit trials at $\ell$.
Put
\[
 U_\ell=\bigcup_{h\in S_\ell}T_h,\qquad |U_\ell|\le DL.
\]
Any failed trial at $\ell$ implies $T_\ell\subseteq U_\ell$,
because $\col_P(h)\subseteq T_h$.

Condition also on all color-tables other than $T_\ell$.
Since $\ell\notin S_\ell$, the set $U_\ell$ is fixed and
$T_\ell$ remains an independent uniform $L$-subset of $P$.
Hence
\begin{equation}
\begin{split}
 \Prb(T_\ell\subseteq U_\ell\mid U_\ell)
 &=\frac{\binom{|U_\ell|}{L}}{\binom KL}\\
 &\le\left(\frac{|U_\ell|}{K}\right)^L
 \le\left(\frac{DL}{K}\right)^L
 =4^{-L}=\frac1k.
\end{split}
\label{eq:subset}
\end{equation}
This bounds the probability of any failure at $\ell$ during
the run.
Sum over the labels and remove the conditioning.
\end{proof}

\subsection{Finishing the proof}\label{sec:count}

We have shown
\begin{equation}
 m_{\mono}\le M/k,\qquad
 \E m_{\highdeg}\le M/k,\qquad
 \E m_{\lowdeg}\le M/k.
 \label{eq:children}
\end{equation}
Now, we are ready to finish the proof.

\begin{proof}[Proof of Theorem~\ref{thm:randomized}]
\textit{Properness.}
Child-mono preserves the recursive input condition by
Lemma~\ref{lem:mono}.
The other children retain parent labels, so their virtual graphs
are subgraphs of the parent's virtual graph.
Restricting to one budget run also preserves the condition.

An accepted CBIP color avoids the accepted colors on the
opposite side of its current component.
Every earlier actual neighbor colored from that anchor's palette
lies on this side.
An accepted FirstFit color avoids the shared-palette colors
of every neighboring group.
At depth zero, the groups are independent and receive private colors.
All other palettes are disjoint.
Induction on the remaining depth therefore proves properness.

\medskip\noindent\textit{Local cost and child budgets.}
Set $A=8kL+q$.
The shared palette has
\[
 K=4kL\ceil{\sqrt M}\le8kL\sqrt M
\]
colors, and the expected number of anchors is at most
$M/\ceil{\sqrt M}\le\sqrt M$.
Thus the expected local color cost is at most
 $A\sqrt M$.

For a fixed child stream with $m\ge1$ labels, its budgets are
$1,2,\ldots,2^s$, where $2^s<2m$.
The geometric sums give
\[
 \sum_{j=0}^s\sqrt{2^j}<5\sqrt m,
 \qquad
 \sum_{j=0}^s2^j<4m.
\]
Consequently, a bound $a_t\sqrt M+d_tM$ for $B_t(M)$ on every
fixed input gives a bound
\begin{equation}
 5a_t\sqrt m+4d_tm
 \label{eq:wrapper}
\end{equation}
for $A_t$.
An empty stream contributes zero.

\medskip\noindent\textit{The recurrence.}
We prove
\[
 \E[\text{colors used by }B_t(M)]
 \le a_t\sqrt M+d_tM,
\]
where
\begin{equation}
\begin{split}
 a_0&=0,\qquad d_0=1,\\
 a_t&=A+\frac{15}{\sqrt k}a_{t-1},
 \qquad
 d_t=\frac{12}{k}d_{t-1}.
\end{split}
\label{eq:recurrence}
\end{equation}
The base case uses at most $M$ private colors.

For the induction step, condition on the parent's random choices.
Its three child streams are then fixed independently of their
fresh random choices, since the parent never uses their output.
Apply~\eqref{eq:wrapper} and average over the parent choices.
With $\mathcal I=\{\mono,\highdeg,\lowdeg\}$, we get
\begin{align*}
 \E[\text{colors used by }B_t(M)]
 &\le A\sqrt M
   +5a_{t-1}\sum_{i\in\mathcal I}\E\sqrt{m_i}
   +4d_{t-1}\sum_{i\in\mathcal I}\E m_i\\
 &\le A\sqrt M
   +15a_{t-1}\sqrt{M/k}
   +12d_{t-1}M/k\\
 &=a_t\sqrt M+d_tM.
\end{align*}
Here~\eqref{eq:children} and concavity give
$\E\sqrt{m_i}\le\sqrt{\E m_i}\le\sqrt{M/k}$.

Since $15/\sqrt k=15/256<1/2$ and $12/k<1/2$,
\[
 a_t=A\sum_{j=0}^{t-1}(15/\sqrt k)^j\le2A,
 \qquad
 d_t=(12/k)^t\le2^{-t}.
\]
At the root,
\[
 \E C_{\mathcal A}(G,\pi)
 \le2A\sqrt n+2^{-T}n
 <2A\sqrt n+1
 =O(\sqrt n).
\]

Each forwarding decreases the remaining depth, so a vertex
visits at most $T+1$ instances.
Every budget is less than $2n$.
Maintaining the virtual graphs, classes, components, and color sets,
and sampling the anchors and color-tables, takes expected
polynomial time.
This proves the theorem.
\end{proof}

\section*{Declaration of AI Usage}

The results and analysis are developed using ChatGPT Astra and the author's previous preliminary studies. 
The author substantially simplified and reorganized  the presentation that the AI tool gives. These includes but not limited to: simplify the proof for the lower bound and the randomized algorithm, write overviews and discuss the underlying ideas and connections to previous studies, sketch illustrations and flowchart that help present the ideas more clearly, etc.  The author has verified the proofs  and take full responsibility for the main contents (Theorem \ref{thm:deterministic} and Theorem \ref{thm:randomized}) of this article. The content in the appendix has only be loosely checked and is subject to future update.

\bibliographystyle{alpha}
\bibliography{references}

\newpage

\appendix

\section{A lower bound against randomized algorithms}
\label{app:random-lower}

An oblivious adversary fixes the ordered input independently of
the algorithm's random choices. We construct such an input
distribution. All its graphs are three-colorable, including
outcomes on which the simulation fails. Designated sets and
bipartitions are part of the construction, not additional
information supplied to the algorithm. All logarithms are to
base $2$, except for $\ln$.

\begin{theorem}\label{thm:randomized-lower}
There is an absolute constant $c>0$ such that, for every sufficiently
large $n$ and every randomized online coloring algorithm
$\mathcal A$, some fixed ordered three-colorable graph satisfies
\[
 |V(G)|=n,\qquad
 \E C_{\mathcal A}(G,\pi)
 \ge 2^{c(\log n)^{1/3}(\log\log n)^{2/3}}.
\]
The expectation is over the algorithm's random choices.
The statement holds also when $n$ is supplied to the algorithm.
\end{theorem}

The constructions below replace the color-dependent choices
in Section~\ref{sec:lower} by input randomness. They always run
to completion. Reaching $K$ colors is an event used in the
analysis, not a reason to stop or change the input.

\subsection{Oblivious bipartite constructions}

\begin{lemma}\label{lem:oblivious-independent}
For integers $t\ge1$ and $j\ge0$, there is a distribution on
ordered bipartite graphs with a fixed number
$m_j\le2\cdot32^j$ of vertices and a designated independent set $I$ such that
every deterministic $t$-fold algorithm satisfies
\[
 \Prb\bigl(|\col(I)|\ge t+j\ceil{t/2}\bigr)\ge\frac12.
\]
A bipartition placing $I$ on one side is determined by the input randomness.
The guarantee holds after any earlier history, also when all
new vertices are joined to a common fixed earlier set $U$.
\end{lemma}
\begin{proof}
At $j=0$, present one vertex and designate it.
For $j\ge1$, present two independent level-$(j-1)$ constructions,
with designated sets $I_1,I_2$, and then a vertex $x$ adjacent
to every vertex of $I_1$.
An independent fair coin designates either
$I_1\cup I_2$ or $I_2\cup\{x\}$.
This coin affects no edge already presented and can remain
hidden until these vertices have been colored.

Repeat this procedure sixteen times on disjoint vertices,
and designate the union of their chosen sets.
Put $s=\ceil{t/2}$ and $a=t+(j-1)s$.
Conditional on any history before a repetition, the two child
designated sets both have at least $a$ colors with probability
at least $1/4$. This follows by applying the induction hypothesis
first to the first child and then to the second child after
the first has been revealed.

On that event, set
\[
 P=\col(I_1),\qquad Q=\col(I_2),\qquad X=\col(\{x\}).
\]
We have $|X|=t$ and $X\cap P=\varnothing$.
Lemma~\ref{lem:union} shows that one candidate designated set
has at least $a+s$ colors. The hidden coin selects such a set
with probability at least $1/2$. Each repetition therefore
succeeds with conditional probability at least $1/8$, and
\[
 \Prb(\text{all sixteen repetitions fail})
 \le(7/8)^{16}<\frac12.
\]

Both choices lie on one side of suitable bipartitions.
For the first choice, put $I_1,I_2$ on the same side and $x$
on the other side. For the second, put $I_2,x$ on the same side,
opposite $I_1$. Orient the sixteen disjoint repetitions with
their designated sets on the same side. These choices depend only
on input coins. Finally,
\[
 m_0=1,\qquad m_j=16(2m_{j-1}+1)\le2\cdot32^j.
\]
Additional adjacency to $U$ only forbids colors, so the same
argument applies. All coins can be sampled before the input starts.
\end{proof}

\begin{lemma}\label{lem:oblivious-two-sided}
Let $K\ge2$ and $1\le t\le q\le K$.
There is an oblivious distribution on ordered bipartite graphs,
with a designated bipartition and a fixed number
\[
 B\le CK\,2^{Cq/t}
\]
of vertices, such that every deterministic $t$-fold algorithm,
after any earlier history, has probability at least $1/9$ of
either using $K$ colors globally or producing $q$ two-sided colors
in this graph. All its vertices may additionally be joined to
a common fixed earlier set $U$.
\end{lemma}
\begin{proof}
Present $4K$ independent copies of
Lemma~\ref{lem:oblivious-independent}, with
\[
 j=\ceil{\frac{8q-t}{\ceil{t/2}}}=O(q/t).
\]
Each designated set has at least $8q$ colors with conditional
probability at least $1/2$.
For analysis, count a copy if this occurs or the global target
$K$ has been reached by its completion. The expected number
of uncounted copies is at most $2K$.
Markov's inequality gives probability at least $1/3$ that at
least $K$ copies are counted.

Suppose this event occurs and the target has not been reached.
Let $p\le K-1$ be the global number of colors.
The incidence graph between any $K$ counted copies and these
$p$ colors has left degree at least $8q$.
Lemma~\ref{lem:degree-count}, with $a=K>p=b$ and $d=8q$,
gives $8q$ colors each appearing in the designated sets of
at least two copies.

Independently flip the structural bipartition of each of the
$4K$ copies by a fair coin. These flips affect no edge within
this construction, so they remain independent of its coloring.
Each repeated color has probability at least $1/2$ of occurring
on both sides of the resulting bipartition.
Fix $8q$ such colors for analysis, and let $X$ count those
that become two-sided. Conditional on the history before the flips,
\[
 \E X\ge4q,\qquad 0\le X\le8q.
\]
Since $\E X\le q+8q\Prb(X\ge q)$, we have
$\Prb(X\ge q)\ge3/8\ge1/3$.
The desired event thus has probability at least $1/9$.
The selected copies and colors are used only for analysis;
they determine no edge. Finally,
\[
 B=4Km_j\le CK\,2^{Cq/t}.
\]
The assertion with extra adjacency to $U$ follows from the
preceding lemma.
\end{proof}

\subsection{An oblivious simulation lemma}

\begin{lemma}\label{lem:oblivious-simulation}
Suppose a distribution on ordered three-colorable graphs has
positive integer size and post-degree bounds $N,D$ and forces
every deterministic $q$-fold algorithm to use $K$ colors with
probability at least $\rho$.
For $1\le t<q\le K$ and $0<\varepsilon<1$, there is an oblivious
distribution forcing every deterministic $t$-fold algorithm
to use $K$ colors with probability at least $\rho-\varepsilon$.
Every graph in its support is three-colorable, and its bounds satisfy
\begin{equation}
 N'\le FND,\qquad D'\le FD,
 \qquad
 F=CK^2\ln\!\left(\frac{2NKD}{\varepsilon}\right)2^{Cq/t}.
 \label{eq:oblivious-simulation}
\end{equation}
\end{lemma}
\begin{proof}
Pad the given distribution to length $N$ with isolated vertices.
For each simulated vertex $v$, let $\sigma(v)$ consist of
\begin{equation}
 R=\ceil{144KD\ln\!\left(\frac{2NKD}{\varepsilon}\right)}
 \label{eq:oblivious-copies}
\end{equation}
independent copies of Lemma~\ref{lem:oblivious-two-sided},
each of the same fixed size $B$.

Independently assign every copy in $\sigma(w)$ a uniform index
from $[D]$, for its possible future neighbor.
When $v$ arrives and is the $a$-th future neighbor of $w$,
select the first designated side of every copy in $\sigma(w)$
having index $a$. Let $U_v$ be the union of these sides over
all earlier neighbors $w$ of $v$. Join every vertex of
$\sigma(v)$ to all of $U_v$.

Each copy serves at most one future simulated vertex.
All edges depend only on the simulated graph and independent
input coins, not on online colors.
The indices assigned within $\sigma(v)$ affect no edge until
a later simulated vertex arrives. For analysis, reveal them
only after $\sigma(v)$ has been colored.

\medskip\noindent\textit{Three-colorability.}
Fix a proper three-coloring $\phi$ of the simulated graph.
Color every copy in $\sigma(w)$ with the two colors other than
$\phi(w)$. If it is used for a later neighbor $v$, orient it
so its selected side has color $\phi(v)$.
There is at most one such requirement per copy, and
$\phi(v)\ne\phi(w)$. Orient unused copies arbitrarily.
Then $U_v$ is monochromatic with color $\phi(v)$, whereas
$\sigma(v)$ uses the other two colors. This gives a proper
three-coloring on every outcome of the input coins.

\medskip\noindent\textit{Repeated colors.}
Fix a deterministic algorithm for the actual $t$-fold input.
For copy $a$ in $\sigma(v)$, let $Y_a$ indicate that it has
$q$ two-sided colors or that the global target has been reached
by its completion. After every preceding history,
\[
 \Prb(Y_a=1\mid\mathcal F_{a-1})\ge\frac19.
\]
Thus
\[
 \E(2^{-Y_a}\mid\mathcal F_{a-1})\le\frac{17}{18},
 \qquad
 \E\,2^{-\sum_{a=1}^R Y_a}\le e^{-R/18}.
\]
Markov's inequality yields
\begin{equation}
 \Prb\!\left(\sum_{a=1}^R Y_a<R/18\right)
 \le 2^{R/18}e^{-R/18}\le e^{-R/72}.
 \label{eq:oblivious-good-copies}
\end{equation}

Suppose fewer than $K$ global colors have appeared and
$G\ge R/18$ copies have $q$ two-sided colors.
Choose $q$ such colors from each good copy for analysis.
At least $q$ distinct colors each occur in at least $G/(2K)$
of these selections. Otherwise their total number of occurrences
would be at most
\[
 (q-1)G+(K-q)\frac{G}{2K}<qG,
\]
contrary to the $qG$ selected occurrences.
Choose $q$ such colors as the provisional simulated set $\col(v)$,
before revealing the indices in $\sigma(v)$.
Each is two-sided in at least $R/(36K)$ copies.

\medskip\noindent\textit{Coverage of future-neighbor positions.}
Conditional on the colored copies and this provisional set,
their indices are still independent and uniform on $[D]$.
For one provisional color and one index, the probability that
no copy carrying that color on both sides receives the index is at most
\[
 (1-1/D)^{R/(36K)}\le e^{-R/(36KD)}.
\]
Except with probability $qD e^{-R/(36KD)}$, every provisional
color therefore occurs on selected sides at every index.

Let $\mathcal E_v$ mean that the actual target has not been
reached by completion of $\sigma(v)$ and either the required
number of good copies or this coverage property fails.
Then, conditional on the preceding history,
\[
 \Prb(\mathcal E_v)
 \le e^{-R/72}+qD e^{-R/(36KD)}\le\frac{\varepsilon}{N}.
\]
Indeed, put $Z=2NKD/\varepsilon$.
By~\eqref{eq:oblivious-copies}, the terms are at most
$Z^{-2}$ and $KD Z^{-4}$, respectively, each at most
$\varepsilon/(2N)$. Consequently,
\begin{equation}
 \Prb\left(\bigcup_v\mathcal E_v\right)\le\varepsilon.
 \label{eq:oblivious-all-defects}
\end{equation}

\medskip\noindent\textit{The simulated coloring.}
Define an online $q$-fold algorithm on the simulated input.
It runs the fixed actual algorithm with the independent input
coins above. Until the actual target is reached or the first
event $\mathcal E_v$ occurs, it supplies the provisional set
$\col(v)$. At either event, it switches permanently to fresh
$q$-sets, disjoint from all previously supplied colors and
from each other.

This algorithm is proper on every outcome. Before the switch,
for any earlier neighbor $w$ of $v$, coverage puts every color
of $\col(w)$ in $U_v$. Every actual vertex of $\sigma(v)$
avoids these colors, as does $\col(v)$. Fresh sets preserve
properness after the switch.

The simulated algorithm uses only the revealed simulated
prefix, and its coins are independent of that input.
Fixing them gives a deterministic $q$-fold algorithm, so the
hypothesis implies simulated success with probability at least
$\rho$. If the actual target is never reached and no
$\mathcal E_v$ occurs, no switch happens and every simulated
color is an actual color. Equation~\eqref{eq:oblivious-all-defects}
therefore gives actual success probability at least
$\rho-\varepsilon$. The provisional sets and the switch
affect only the analysis, never the actual input distribution.

\medskip\noindent\textit{Size and post-degree.}
We have $N'=NRB$. A vertex has at most $B-1$ future neighbors
inside its own copy. Its external future neighbors belong to
the unique later collection served by that copy. Hence
\[
 D'\le B-1+RB<(R+1)B.
\]
Together with $B\le CK2^{Cq/t}$ and
\eqref{eq:oblivious-copies}, this proves
\eqref{eq:oblivious-simulation}.
\end{proof}

\subsection{Finishing the proof}

\begin{proof}[Proof of Theorem~\ref{thm:randomized-lower}]
Fix a sufficiently large target $K$, and set
\[
 \ell=\log K,\quad \beta=\ceil{\ell},\quad
 h=\ceil{\log_\beta K},\quad t_h=K,\quad
 t_i=\beta^i\quad(0\le i<h).
\]
Then $h=O(\ell/\log\ell)$ and $t_{i+1}/t_i\le\beta$.
A singleton in the $K$-fold model reaches the target with
probability one, with $N_h=D_h=1$.
Apply Lemma~\ref{lem:oblivious-simulation} successively with
\[
 q=t_{i+1},\qquad t=t_i,\qquad \varepsilon=\frac1{4h},
 \qquad i=h-1,\ldots,0.
\]
The final ordinary-coloring distribution forces $K$ colors
against every deterministic algorithm with probability at least $3/4$.

We check that the additional logarithmic factor does not change
the bound. As long as $N,D\le2^{\ell^4}$,
\[
 CK^2\ln(8hNKD)\,2^{C\beta}\le K^{C_0}
\]
for an absolute constant $C_0$ and sufficiently large $K$.
Put $F=K^{C_0}$. The size and post-degree recurrences give,
as in~\eqref{eq:lower-induction},
\[
 D_i\le F^{h-i},\qquad
 N_i\le F^{(h-i)(h-i+1)/2}.
\]
These bounds stay inside the assumed envelope, since
\[
 \log N_i,\log D_i
 \le O(\ell h^2)
 =O\!\left(\frac{\ell^3}{(\log\ell)^2}\right)<\ell^4.
\]
Induction therefore justifies the envelope and gives
\[
 N_0\le2^{\,O((\log K)^3/(\log\log K)^2)}.
\]

For each fixed seed of a randomized algorithm, this same input
distribution forces $K$ colors with probability at least $3/4$.
Averaging over its independent seed gives expected color count
at least $3K/4$, averaged over the input distribution.
Some fixed ordered graph in the support consequently has
expected color count at least $3K/4$.

Choose
\[
 K=\ceil{2^{2c(\log n)^{1/3}(\log\log n)^{2/3}}}
\]
for a sufficiently small absolute constant $c>0$.
The same inversion as in the deterministic proof gives
$N_0\le n$ and, for sufficiently large $n$,
\[
 \frac{3K}{4}\ge
 2^{c(\log n)^{1/3}(\log\log n)^{2/3}}.
\]
Pad to exactly $n$ vertices with isolated vertices.
All distributional guarantees hold for every deterministic
algorithm, including one supplied with this value of $n$.
\end{proof}

\section{A deterministic algorithm in the random order model}
\label{app:rom}

In the random order model, the graph is fixed and its vertices arrive
in a uniformly random order.
The algorithm sees only the edges to earlier vertices.
We do not assume fixed vertex identifiers: arrival positions are the
only names available to the algorithm.

\begin{theorem}\label{app:rom-theorem}
There is a deterministic polynomial-time online coloring algorithm
$\mathcal D$ that properly colors every ordered three-colorable graph
and, on every fixed three-colorable graph $G$ with $n$ vertices,
satisfies
\[
 \E_\pi C_{\mathcal D}(G,\pi)=O(\sqrt n),
\]
where $\pi$ is uniform.
The algorithm uses no external random bits and need not know $n$.
\end{theorem}

We first implement a variant of the randomized algorithm with
$O(\log^2 n)$ fair bits.
We then obtain those bits from already revealed edges, using few
colors before starting that algorithm.

\subsection{A finite random seed}\label{app:rom-finite-seed}

The following variant keeps the same virtual graphs, color sets,
and routing rules as Section~\ref{sec:algorithm}.
It changes the anchor probabilities, the shared-palette size,
and the distribution of the color-tables.
Every color-table still has $L$ distinct colors.

\begin{lemma}\label{app:rom-seed}
For every integer $N\ge1$, there is a seeded online algorithm
$\mathcal A_N$ using
\[
 R(N)=8\bigl(32+\ceil{\log(N+1)}\bigr)\ceil{\log(N+1)}
\]
fair bits, such that every seed gives a proper coloring and every
fixed ordered three-colorable input with at most $N$ vertices has
expected color cost at most $C_0\sqrt N$.
Here $C_0$ is absolute, and the running time is polynomial in $N$.
\end{lemma}
\begin{proof}
Keep $k=2^{16}$, $L=8$, and $q=36$.
For a run with label budget $M$, use
\[
 r=2^{\ceil{\log\sqrt M}},\qquad
 D=kr,\qquad K=2^{10}DL,
\]
sampling anchors with probability $1/r$.
These quantities are powers of two.
Use the root $B_T(N)$ with $T=\ceil{\log(N+1)}$ and the same
child-budget procedure as in the main algorithm.

We describe sufficient local independence before generating it from
the seed.
Pairwise independent anchor indicators suffice.
Indeed, for the fixed neighbor set $S$ at a label's first request
of degree greater than $D$, let $Y$ count sampled neighbors.
Then
\[
 \lambda=\E Y=|S|/r>k,\qquad
 \operatorname{Var}(Y)\le\lambda.
\]
The argument of Lemma~\ref{lem:high} therefore gives
\[
 \Prb(\ell\text{ enters child-highdeg})
 \le\Prb(Y=0)\le1/\lambda<1/k.
\]

Identify the ordered shared palette with the field $\mathbb F_K$.
Fix distinct elements $\eta_1,\ldots,\eta_L$.
For each label $\ell$, obtain uniform field parameters
$a_\ell,b_\ell$, put
\[
 a'_\ell=
 \begin{cases}
  a_\ell,&a_\ell\ne0,\\
  1,&a_\ell=0,
 \end{cases}
 \qquad
 T_\ell=\{a'_\ell\eta_j+b_\ell:1\le j\le L\}.
\]
The table has $L$ distinct colors.
Require six-wise independence of all parameter positions
$(\ell,0),(\ell,1)$ and independence from the anchor choices.

For any two prescribed values, the probability that the first two
displayed table entries equal those values is at most $2/K^2$.
Indeed, the pair determines $a'_\ell,b_\ell$, and this parameter
pair has at most two preimages among the $K^2$ equally likely
choices of $a_\ell,b_\ell$.

Condition on the anchors and use the last eligible neighbor set
$S_\ell$ from Lemma~\ref{lem:low}.
A failed FirstFit trial implies that each of the first two table
entries agrees with some entry at a label in $S_\ell$.
There are at most $(DL)^2$ specifications of these agreements.
A fixed specification involves at most three labels and six
parameter positions.
Their joint independence and the preceding bound give
\[
 \Prb(\ell\text{ ever enters child-lowdeg})
 \le 2(DL/K)^2=2^{-19}\le1/k.
\]
This argument conditions only on the parameters occurring in a
specified agreement, rather than on all other color-tables.

The child-mono bound from Corollary~\ref{cor:overflow} and
Lemma~\ref{lem:mono} is deterministic and unchanged.
Here the proof of Lemma~\ref{lem:growth} applies also to retained
color sets, repeated requests, and later edge additions:
the occurrences in its argument are accepted requests,
components only merge, and a retained color on the opposite side
forbids a new occurrence.
For a failed trial, apply that argument after hypothetically
accepting its color; no earlier color set changes.
The expected local cost is at most $A_*\sqrt M$, where
\[
 A_*=2^{11}kL+q.
\]
Thus the recurrence~\eqref{eq:recurrence}, with $A_*$ in place of
$A$, again gives a root cost at most $(2A_*+1)\sqrt N$.

It remains to generate the choices with a short seed.
Put $w=32+\ceil{\log(N+1)}$ and fix a binary representation of
$\mathbb F_{2^w}$.
All descendant budgets are below $2N$.
Consequently all local first-appearance positions and their two
parameter positions have injective field encodings, and
\[
 K<2^{30}\sqrt{2N}<2^{32}(N+1)\le2^w.
\]

At each recursion distance $d=0,\ldots,T-1$, choose independently
a uniform polynomial of degree at most one and an independent
uniform polynomial of degree at most five over $\mathbb F_{2^w}$.
Use the first polynomial at a label's local first-appearance position;
the label is an anchor when the first $\log r$ binary coordinates
of its value are zero.
Use the second polynomial at the two parameter positions of each
label; project its values onto their first $\log K$ coordinates
and interpret these as elements of $\mathbb F_K$.
Polynomial evaluations give the required pairwise and six-wise
independence, respectively, by the Vandermonde formula.

The polynomial pair at distance $d$ is shared by all runs at that
distance.
To justify this reuse, condition on the polynomial choices at smaller
distances.
All streams, labels, and budget boundaries at distance $d$ are then
fixed: parents never use their children's output.
The current polynomial pair is independent of that conditioning,
so every run has the local marginal distributions just analyzed.
Sharing may correlate different runs, but their costs are added
in expectation.
After exposing this distance's pair, the same argument applies to
the next distance.
Thus the main inductive color-count proof remains valid.

There are eight field coefficients at each of $T$ distances,
requiring exactly $8wT$ fair bits.
All further operations are deterministic.
The fields and their representations can be constructed in polynomial
time in $N$; even enumerating binary polynomials and testing
irreducibility suffices, since $2^w=O(N+1)$.
Properness uses only the retained color sets and disjoint palettes,
so it holds for every seed.
\end{proof}

\subsection{Obtaining bits from revealed edges}\label{app:rom-bits-section}

Fix a desired seed length $R\ge1$.
For a completed block $(v_1,\ldots,v_a)$, put $p=\floor{a/3}$.
Pair its first $2p$ positions consecutively, and let
\[
 W=(v_{2p+1},\ldots,v_a)
\]
be the ordered witness list.
The signature of a paired vertex $x$ is
\[
 s_W(x)=
 \bigl(\mathbf1_{xv_{2p+1}\in E},\ldots,
       \mathbf1_{xv_a\in E}\bigr).
\]
A pair is usable if its signatures differ.
A usable ordered pair $(x,y)$ supplies the bit
$\mathbf1_{s_W(x)>_{\mathrm{lex}}s_W(y)}$.
The block test succeeds when there are at least $R$ usable pairs,
and outputs the first $R$ bits in pair-position order.
The test is performed only after the block is complete, when all
these edges have been revealed.

\begin{lemma}\label{app:rom-bits}
For a fixed block of positions in a uniform random presentation,
condition on the unordered two-element set in each pair, the ordered
witness list, and the entire order outside the block.
Usability is then fixed, and the usable pairs supply independent
unbiased bits.
\end{lemma}
\begin{proof}
The remaining possibilities are the independent orientations of the
paired sets, all equally likely.
Reversing a usable pair reverses its bit and changes neither its
usability nor the indices selected by the test.
\end{proof}

We next bound the cost of waiting for a successful block.
Write $\nu(F)$ for the maximum matching size of a graph $F$.
Ordinary FirstFit uses at most $2\nu(F)+1$ colors in every order.
Indeed, for each $i\le\floor{c/2}$, a vertex of FirstFit color $2i$
has an earlier neighbor of color $2i-1$.
If $c$ colors are used, these edges form a matching because their
pairs of endpoint colors are disjoint.

\begin{lemma}\label{app:rom-detection}
Let $F$ be a fixed three-colorable graph on $m$ vertices, with
$\nu=\nu(F)\ge1024$.
In a uniform random order, test a fixed interval of $a\ge m/4$
positions.
If $Q$ is its number of usable pairs, then
\[
 \Prb(Q<\alpha\nu)\le e^{-\beta\nu},
 \qquad
 \alpha=2^{-22},\quad\beta=2^{-23}.
\]
\end{lemma}
\begin{proof}
First find $\rho=\floor{\nu/8}$ disjoint edges $x_iw_i$ and a set
$Z$, disjoint from their endpoints, satisfying
\[
 |Z|\ge m/5,\qquad E(\{w_1,\ldots,w_\rho\},Z)=\varnothing.
\]
For this purpose only, fix an offline independent color class
$I$ of size at least $m/3$.
Let $h$ be the maximum matching size between $I$ and its complement.

If $h\ge\nu/8$, choose $\rho$ crossing edges with $w_i\in I$
and set $Z=I\setminus\{w_1,\ldots,w_\rho\}$.
Then $|Z|\ge m/3-m/16>m/5$.
Otherwise, let $U$ be the endpoints of a maximum crossing matching.
These endpoints cover every crossing edge, and $|U|<\nu/4$.
Deleting $U$ from a maximum matching of $F$ leaves more than
$3\nu/4$ edges, all outside $I\cup U$.
Choose $\rho$ of them and set $Z=I\setminus U$.
Now
\[
 |Z|\ge m/3-\nu/4\ge5m/24>m/5,
\]
and the required nonadjacencies hold.

View the random order as a uniform placement into $m$ positions.
The tested block has at least $m/12$ witness positions and at
least $m/6-2$ paired positions.
Expose the placement in $s=\floor{\rho/6}$ trials.
At each trial choose an edge whose endpoints remain unexposed.
Reveal the positions of $w_i,x_i$ and, if $x_i$ occupies a paired
position whose partner is unexposed, reveal that partner.

A trial is good if $w_i$ occupies a witness position, $x_i$ occupies
such an intact paired position, and its partner lies in $Z$.
Each trial exposes at most three vertices and touches at most three
of the chosen disjoint edges, so an untouched edge is always available.
Before a trial fewer than $\rho/2\le m/32$ vertices have been exposed.
Conditional on the past, the remaining placement is uniform.
The three successive favorable probabilities are at least
\[
 \frac{m/12-m/32}{m}\ge\frac1{32},
 \qquad
 \frac{m/6-2-2(m/32+1)}m\ge\frac1{16},
 \qquad
 \frac{m/5-m/32-2}{m}\ge\frac18.
\]
Here $m\ge2\nu\ge2048$.
Thus every trial is conditionally good with probability at least
$p_0=2^{-12}$.

A good trial gives a usable pair, because $x_i$ is adjacent to the
witness $w_i$, whereas its partner in $Z$ is not.
Different good trials give different pairs.
If $S$ counts good trials, then $Q\ge S$.
Successive conditioning gives, for $\lambda>0$,
\[
 \E e^{-\lambda S}
 \le(1-p_0+p_0e^{-\lambda})^s.
\]
Taking $\lambda=\ln2$ and applying exponential Markov yields
$\Prb(S<p_0s/2)\le e^{-p_0s/8}$.
Finally,
\[
 \rho\ge\nu/16,\qquad
 s\ge\rho/12\ge\nu/192,\qquad
 p_0s\ge\nu/2^{20}.
\]
The stated $\alpha,\beta$ therefore suffice.
\end{proof}

\subsection{The deterministic algorithm}\label{app:rom-conversion-section}

\begin{lemma}\label{app:rom-conversion}
Suppose a seeded online algorithm properly colors every ordered
three-colorable graph of at most $N$ vertices for every seed in
$\{0,1\}^R$, and its expected cost under a uniform seed is at most $F$.
There is a deterministic online algorithm with expected cost
\[
 F+O\bigl(R+\log\log(N+2)\bigr)
\]
in random order.
It needs no fixed vertex identifiers.
\end{lemma}
\begin{proof}
Run ordinary FirstFit while dividing arriving positions into blocks
of lengths $1,2,4,\ldots$.
Test each completed block.
At the first success, use its $R$ bits to run the seeded algorithm
on the remaining induced graph, ignoring the prefix and using a
disjoint palette.
If no block succeeds, FirstFit colors the entire input.
Every order receives a proper coloring.

Let the actual graph have $n\le N$ vertices.
For analysis define the test outcomes of all completed blocks,
even if the algorithm has already switched.
Let $J$ be the first successful block, or the final occupied block
if none succeeds.
Let $P_j$ be the prefix through block $j$, including a final
incomplete block, and put
\[
 X=\nu(G[P_J]),\qquad h=1+\ceil{\log(n+1)}.
\]
For $j\ge1$, the preceding block is complete and occupies more than
$|P_j|/4$ positions.
Conditional on the unordered set $P_j$, its order is uniform.
For $x\ge x_0=\max\{1024,R/\alpha\}$,
Lemma~\ref{app:rom-detection} therefore gives
\[
 \Prb(Q_{j-1}<R,\ \nu(G[P_j])\ge x)\le e^{-\beta x}.
\]
The event $J=j\ge1$ requires $Q_{j-1}<R$.
For $J=0$, the prefix has one vertex and $X=0$.
A union bound gives
\[
 \Prb(X\ge x)\le\min\{1,h e^{-\beta x}\}
 \qquad(x\ge x_0).
\]
Summing the tail yields
\[
 \E X
 \le x_0+\frac{\ln h+1}{\beta}+1
 =O\bigl(R+\log\log(n+2)\bigr).
\]
The FirstFit prefix uses at most $2X+1$ colors.

To analyze the suffix, suppose block $j$ is the first successful one.
Condition on the entire order outside this block, its unordered
paired sets, and its ordered witness list.
All earlier test outcomes and all current usability decisions are
fixed by this information.
The event of first switching here is therefore invariant under
independent reversals of the current pairs.
By Lemma~\ref{app:rom-bits}, the extracted seed remains uniform
under this conditioning, while the ordered suffix is fixed.
The suffix algorithm ignores the prefix, so its conditional expected
cost is at most $F$.
Adding the prefix bound proves the claim.
\end{proof}

\begin{proof}[Proof of Theorem~\ref{app:rom-theorem}]
For a known budget $N$, apply Lemma~\ref{app:rom-conversion} to
$\mathcal A_N$ from Lemma~\ref{app:rom-seed}.
Since $R(N)=O(\log^2(N+1))$, the resulting expected cost is
\[
 C_0\sqrt N+
 O\bigl(\log^2(N+1)+\log\log(N+2)\bigr)
 =O(\sqrt N).
\]

For unknown $n$, divide arriving positions into fixed outer epochs
of capacities $1,2,4,\ldots$.
Run the budget-$N$ algorithm in an epoch of capacity $N$, with
fresh state and a palette disjoint from other epochs.
Conditional on the vertex set of an epoch, its internal order is
uniform, including the incomplete final epoch.
Thus its expected cost is $O(\sqrt N)$.
The largest capacity used is at most $n$, and the geometric sum
of their square roots is $O(\sqrt n)$.

All tests compare adjacency vectors of already revealed vertices;
no fixed vertex identifiers or external random bits are used.
The tests and the seeded algorithm run in polynomial time.
Disjoint palettes at switches and epoch boundaries ensure properness
for every order.
\end{proof}

\end{document}